\documentclass{aa}
\usepackage{epsfig}
\usepackage{graphicx}
\usepackage{amssymb}
\usepackage[below]{placeins}
\usepackage{natbib}
\bibpunct{(}{)}{;}{a}{}{,}

\def \arcmin{$^{\prime}$}

\newcommand{\ICM}{_{\mathrm{ICM}}}
\newcommand{\Kpc}{\,\textrm{kpc}}
\newcommand{\Myr}{\,\textrm{Myr}}

\newcommand{\Kms}{\,\textrm{km}\,\textrm{s}^{-1}}
\newcommand{\gccm}{\,\textrm{g}\,\textrm{cm}^{-3}}
\newcommand{\Jet}{_{\mathrm{jet}}}
\newcommand{\Mpc}{\,\textrm{Mpc}}
\newcommand{\K}{\,\textrm{K}}
\newcommand{\Break}{_{\mathrm{break}}}

\begin{document}

\title{The large-scale shock in the cluster of galaxies Hydra~A}
\author{A. Simionescu\inst{1}
\and E. Roediger \inst{2}
\and P. E. J. Nulsen \inst{3}
\and M. Br\"uggen \inst{2}
\and W. R. Forman \inst{3}
\and H. B\"ohringer\inst{1}
\and N. Werner\inst{4}
\and A. Finoguenov \inst{1,5} 
}

\institute{     Max-Planck-Institute for Extraterrestrial Physics, Giessenbachstr, 85748, Garching, Germany
\and  Jacobs University Bremen, Campus Ring 1, 28759 Bremen, Germany	
\and Harvard-Smithsonian Center for Astrophysics, 60 Garden St., Cambridge, MA 02138, USA	
\and	SRON Netherlands Institute for Space Research, Sorbonnelaan 2, NL - 3584 CA Utrecht, the Netherlands $^6$
\and University of Maryland Baltimore County, 1000 Hilltop Circle, Baltimore, MD, 21250, USA 
 }

\date{Received, accepted }

\abstract{

We analyzed a deep XMM-Newton observation of the cluster of galaxies Hydra~A, focusing on the large-scale shock discovered in Chandra images as a discontinuity in the surface brightness. 
The shock front can be seen both in the pressure map and in temperature profiles in several sectors. 
We compared the results of a spherically symmetric hydrodynamic model to surface brightness profiles and temperature jumps across the shock to determine the shock properties. The Mach numbers determined from the temperature jumps are in good agreement with the Mach numbers derived from EPIC/pn surface brightness profiles and previously from Chandra data and are consistent with M$\sim$1.3. In this simple model, the estimated shock age in the different sectors ranges between 130 and 230 Myr and the outburst energy between 1.5 and $3\times10^{61}$ ergs.
The shape of the shock seen in the pressure map can be approximated with an ellipse centered $\sim$70 kpc towards the NE from the cluster center. This is a good simple approximation to the shock shape seen in the Chandra image, although this shape shows additional small deviations from ellipticity. We aimed to develop a better model that can explain the offset between the shock center and the AGN, as well as give a consistent result on the shock age and energy. To this end, we performed 3D hydrodynamical simulations in which the shock is produced by a symmetrical pair of AGN jets launched in a spherical galaxy cluster. As an explanation of the observed offset between the shock center and the AGN, we consider large-scale bulk flows in the intracluster medium, which were included in the simulation. The simulation successfully reproduces the size, ellipticity, and average Mach number of the observed shock front. The predicted age of the shock is 160 Myr and the total input energy $3\times 10^{61}$ erg. Both values are within the range determined by the spherically symmetric model. To match the observed 70 kpc offset of the shock ellipse from the cluster center by large-scale coherent motions, these would need to have a high velocity of 670$\Kms$. We discuss the feasibility of this scenario and offer alternative ways to produce the observed offset and to further improve the simulation.

}
\maketitle

\footnotetext[6]{now a Chandra fellow at the Kavli Institute for Particle Astrophysics and Cosmology, Stanford University, 382 Via Pueblo Mall, Stanford, CA 94305-4060, USA}

\section{Introduction}

Many clusters of galaxies show central surface brightness peaks associated with a decrease in the temperature of the X-ray emitting gas. In the centers of these clusters, the cooling time usually falls below the Hubble time. Thus, it was initially believed that the X-ray gas in the cluster cores radiates its energy away and, in the absence of other heat sources, cools out of the X-ray band. This was known as the cooling-flow scenario \citep{FabianNulsen77,CowieBinney77}. Early results of the latest generation of X-ray observatories, however, showed that the central gas in cooling-flow clusters does not cool below about one third of the cluster's virial temperature \citep{Peterson01,Peterson03}. This requires a fine-tuned heat source that allows the gas to cool down to 1/3 of its initial temperature but not below this \citep[e.g.][]{Boehringer02}.

Currently, AGN are considered the best candidate as a heating engine to solve the cooling flow problem because they are present in a large fraction of the central galaxies in cool core clusters \citep[71\% of such galaxies are radio loud,][]{Burns90}, and substructures associated with AGN-ICM interaction are frequently observed. The most common among these features are X-ray cavities (also referred to as bubbles) often associated with low-frequency radio emission, which are thought to arise when the AGN injects relativistic plasma into the cluster medium during an outburst. As these bubbles rise buoyantly through the cluster, they deposit their energetic content into the ICM, thus heating it. In general, the most powerful cavities are found in the most X-ray-luminous systems, which require the greatest heat inputs \citep{Birzan04,Rafferty06}. By heating or otherwise disturbing the gas, outbursts can affect the accretion rate, creating a feedback loop that could regulate the energy input and explain the fine-tuning. In some clusters, additional substructure such as X-ray bright filaments \citep[a spectacular example is M87 in the Virgo cluster, e.g.][]{Forman06} and shocks associated with AGN outbursts are also seen, providing several other possible mechanisms by which AGN heat the ICM. 

The cluster of galaxies Hydra A shows a wealth of substructure associated with AGN-ICM interaction. A pronounced set of X-ray cavities was seen in an early Chandra observation \citep{McNamara00,Nulsen02}, and deeper data revealed the presence of a second, larger set of cavities, a large-scale shock and an X-ray bright filament near the cluster center \citep{Nulsen05,Wise07}. Modeling the surface brightness profiles across the shock, \citet{Nulsen05} found Mach numbers between M$\approx$1.2 in the W and M$\approx$1.34 on the NE side and a total energy associated with creating the shock on the order of $10^{61}$ ergs, in agreement with the energy needed to generate the observed cavity system. This ranks the outburst in Hydra A among the most powerful ones known. 

So far, however, a temperature jump associated with the large-scale shock in Hydra A has not been detected. A new exposure with XMM-Newton allows us to find this temperature jump and confirm that classical shock jump conditions apply: both the surface brightness and temperature jumps reflect the same shock Mach number, which is around M$\sim$1.3. We furthermore extend the one-dimensional shock analysis of \citet{Nulsen05} using 3D hydrodynamic simulations, in order to improve our understanding of the shock geometry and the morphology of the cavity system.

\section{Observation and data analysis}

Hydra A was observed with XMM-Newton on December 8th, 2000, for 32.6 kiloseconds (ks) and on May 11th, 2007, for 123 ks. Since the second observation is significantly deeper and large parts of the first observation were affected by soft proton flares, we will focus primarily on the second observation.  

We extracted a lightcurve for each of the three detectors separately and excluded the time periods in the observation when the count rate deviated from the mean by more than 3$\sigma$ in order to remove flaring from soft protons \citep{Pratt02}. After this cleaning, the net effective exposure is $\sim$62 ks for pn, $\sim$81 ks for MOS1, and $\sim$ 85 ks for MOS2. We furthermore excluded CCD 5 of MOS2 from our analysis due to its anomalously high flux in the soft band during the observation \citep[see ][]{snowden2007}.
For data reduction we used the 7.1.0 version of the XMM-Newton Science Analysis System (SAS); the standard analysis methods using this software are described in e.g. \citet{Watson01}.  Out-of-time events were subtracted from the PN data using the standard SAS prescription for the extended full frame mode.

For the background subtraction, we used a combination of blank-sky maps from which point sources have been excised \citep{ReadPonman,Carter07} and closed-filter observations. 
The exact procedure and a short discussion about the agreement between the blank-sky maps and the local background can be found in \citet{Simionescu_HydraAI}. 

\section{Spectral modeling}
We use the SPEX package \citep{kaastra1996} to model our spectra with a plasma model in collisional ionization equilibrium (MEKAL). Unless otherwise stated, the Galactic absorption column density was fixed to $N_{\mathrm{H}}=4.8\times10^{20}$~cm$^{-2}$, the average value from the two available \ion{H}{i} surveys: the Leiden/Argentine/Bonn (LAB) Survey of Galactic \ion{H}{i} \citep[][ $N_{\mathrm{H}}=4.68\times10^{20}$~cm$^{-2}$]{kalberla2005} and the \ion{H}{i} data by \citet{dickey1990} ($N_{\mathrm{H}}=4.90\times10^{20}$~cm$^{-2}$). Point sources identified using the X-ray images are excluded from the spectral analysis. The spectra obtained by MOS1, MOS2 and pn are fitted simultaneously with their relative normalisations left as free parameters. 

The spectra are binned with a minimum of 30 counts per bin and fitted in the 0.35--7 keV band with a Gaussian differential emission measure distribution model ({\it gdem}), which \citet{Simionescu_HydraAI} showed to be the most appropriate for the data. We note that the average temperature is the same for the {\it gdem} model as for a single temperature fit, but the {\it gdem} model usually provides a significantly improved fit. 

Due to the high photon statistics of this deep observation, our best fit reduced $\chi^{2}s$ are sensitive to calibration problems and to the differences between the individual EPIC detectors. To account for this, we include 3\% systematic errors over the entire energy band used for fitting. 

\section{The shock geometry}\label{sect:geom}

\begin{figure*}[!t]
\begin{minipage}{0.48\textwidth}
\includegraphics[width=\columnwidth]{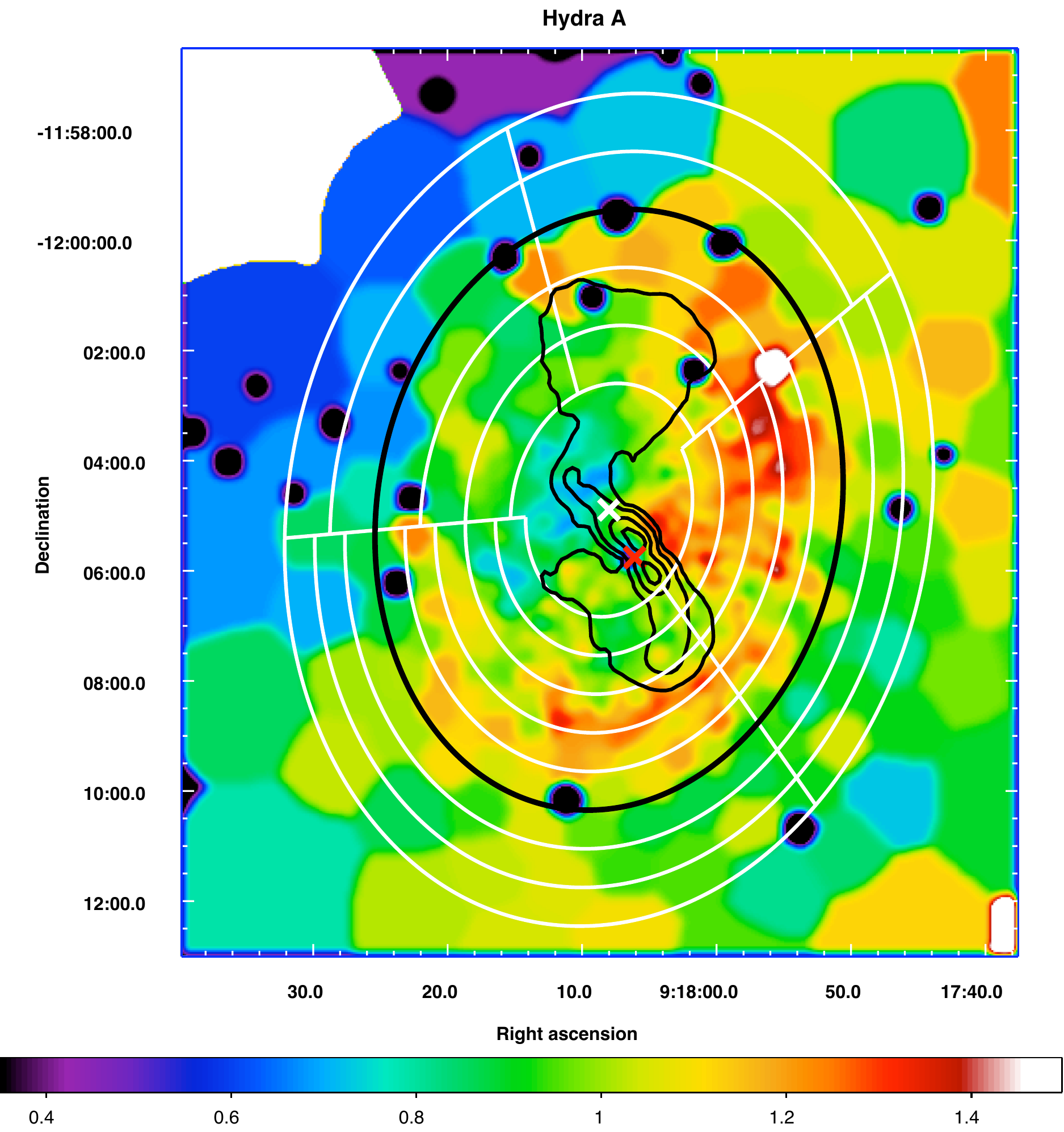}
\caption{Pressure map divided by a radially symmetric, smooth model. 90 cm radio contours \citep{Lane04} are shown in black. The shock is clearly seen as a ring of enhanced pressure, whose shape is simplest approximated by the black ellipse. The center of this ellipse (white cross) is shifted towards the NE with respect to the cluster center (red cross).}
\label{fig:pmap}
\end{minipage}
\hspace{5mm}
\begin{minipage}{0.48\textwidth}
\includegraphics[width=0.9\columnwidth]{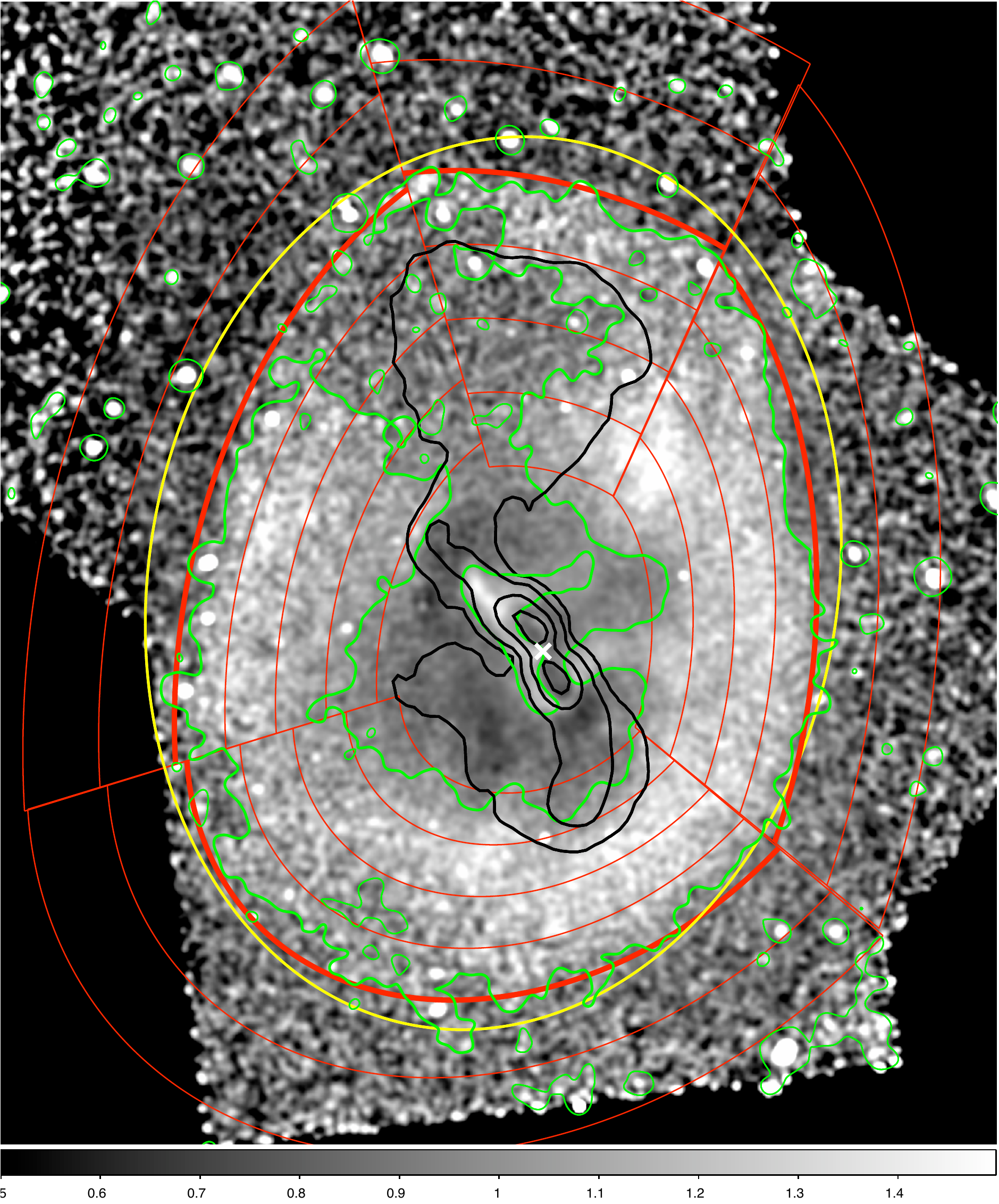}
\caption{Chandra surface brightness map divided by a 2D elliptical beta model. Contours of the surface brightness to beta model ratio are shown in green, 90 cm radio contours \citep{Lane04} are shown in black. The yellow ellipse is the best-fit ellipse to the shock front in the pressure map. The thick red elliptical sectors show a further improved approximation to the shock shape. The white cross marks the cluster center.}
\label{fig:chandrashock}
\end{minipage}
\end{figure*}

Using the fit results for the 2D temperature map of Hydra~A presented in \citet{Simionescu_HydraAI}, we created a quasi-deprojected pressure map of the cluster, with pressure defined as $p = n_{\rm e}kT$. The electron density $n_e$ is determined from the spectrum normalization $Y\propto n_{\rm e}^2 V$, with $V$ being the volume along the line of sight (LOS) corresponding to each spatial bin in the 2D map. Let $R_{\mathrm{max}}$ and $R_{\mathrm{min}}$ be the maximum and minimum radii with respect to the cluster center between which a spatial bin is contained. Assuming that only the gas between the sphere with the minimum radius and the sphere with the maximum radius contribute to the emission \citep[]{Henry04,Mahdavi05} yields, where $L$ is the length of the contributing volume along the LOS and $S$ is the area of the bin in the plane of the sky,
\begin{equation}
L=2\sqrt{(R_{\mathrm{max}}^2-R_{\mathrm{min}}^2)}\\{\rm{and}}\\
V=2SL/3.
\label{eq:alexislos}
\end{equation}
We thus take into account an approximate estimation of the three-dimensional extent of each bin, but assume a constant temperature along the line of sight. Since most of the emission in each bin originates from the densest gas which is found at the smallest effective 3D radii, the map mostly reflects the distribution of the gas pressure in a two dimensional slice through the middle of the cluster, perpendicular to the line of sight. We fitted the radial pressure profile using non-parametric, locally weighted, linear regression smoothing and subsequently divided the pressure map by the resulting radial model in order to reveal small-scale fluctuations. The result is shown in Fig. \ref{fig:pmap}. 

The shock is clearly seen as an approximately elliptical ring of enhanced pressure, on average 20\% higher than the value in the radially smooth model. The simplest approximation to the shock shape seen in the pressure map is an ellipse, indicated in black in Fig. \ref{fig:pmap}, with a semi-major axis of 5.5\arcmin\ (360 kpc) oriented 10 degrees clockwise from the N--S direction, and a semi-minor axis of 4.2\arcmin\ (275 kpc). The center of this ellipse (white cross in Fig. \ref{fig:pmap}) is shifted by $\sim$1.1\arcmin\ (70 kpc) towards the NE with respect to the cluster center (shown with a red cross in the same figure).

We divided the exposure and background corrected Chandra image obtained by combining the two deepest exposures of Hydra~A (totaling 200ks observing time) by the best-fit 2D elliptical beta model (with ellipticity 0.17 and position angle 50.7 degrees counterclockwise from west) in order to highlight the position of the shock front and compare it to the ellipse described above. The resulting map is shown in Fig. \ref{fig:chandrashock}. We find that, although the shock front shape is not exactly elliptical, the ellipse which best matches the shock in the pressure map does provide a good simple approximation, especially taking into consideration that the spatial bins used for the pressure map are up to 0.8\arcmin\ in diameter around the position of the shock. In a more exact approximation, the shock front shape can be described as a set of 4 elliptical sectors, each centered on the cluster center, over-plotted in thick red lines Fig. \ref{fig:chandrashock}.

\section{1D shock model and Mach number estimates}\label{sect:mach}

\begin{figure*}[!ht]
\begin{minipage}{0.5\textwidth}
\includegraphics[width=\columnwidth]{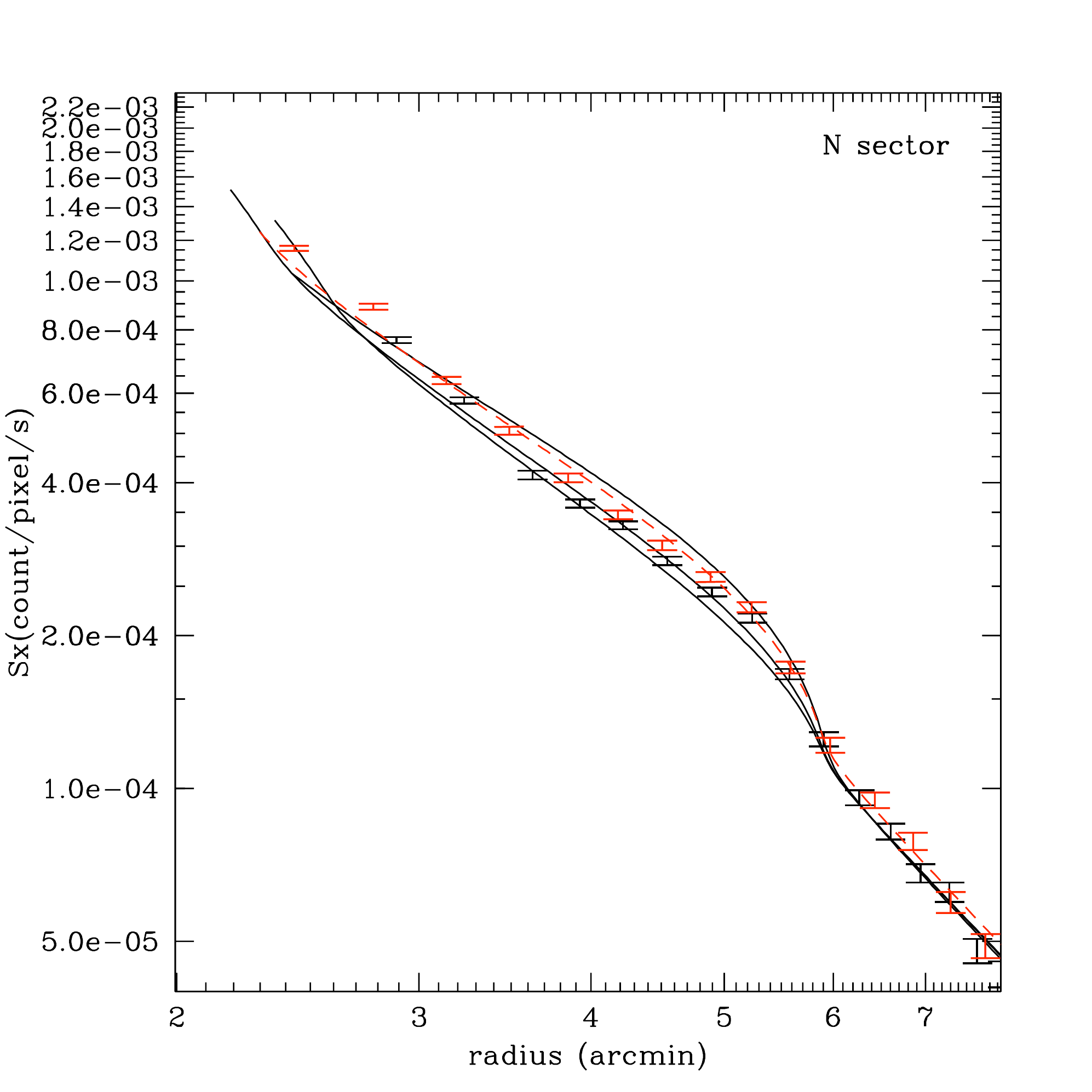}
\end{minipage}
\begin{minipage}{0.5\textwidth}
\includegraphics[width=\columnwidth]{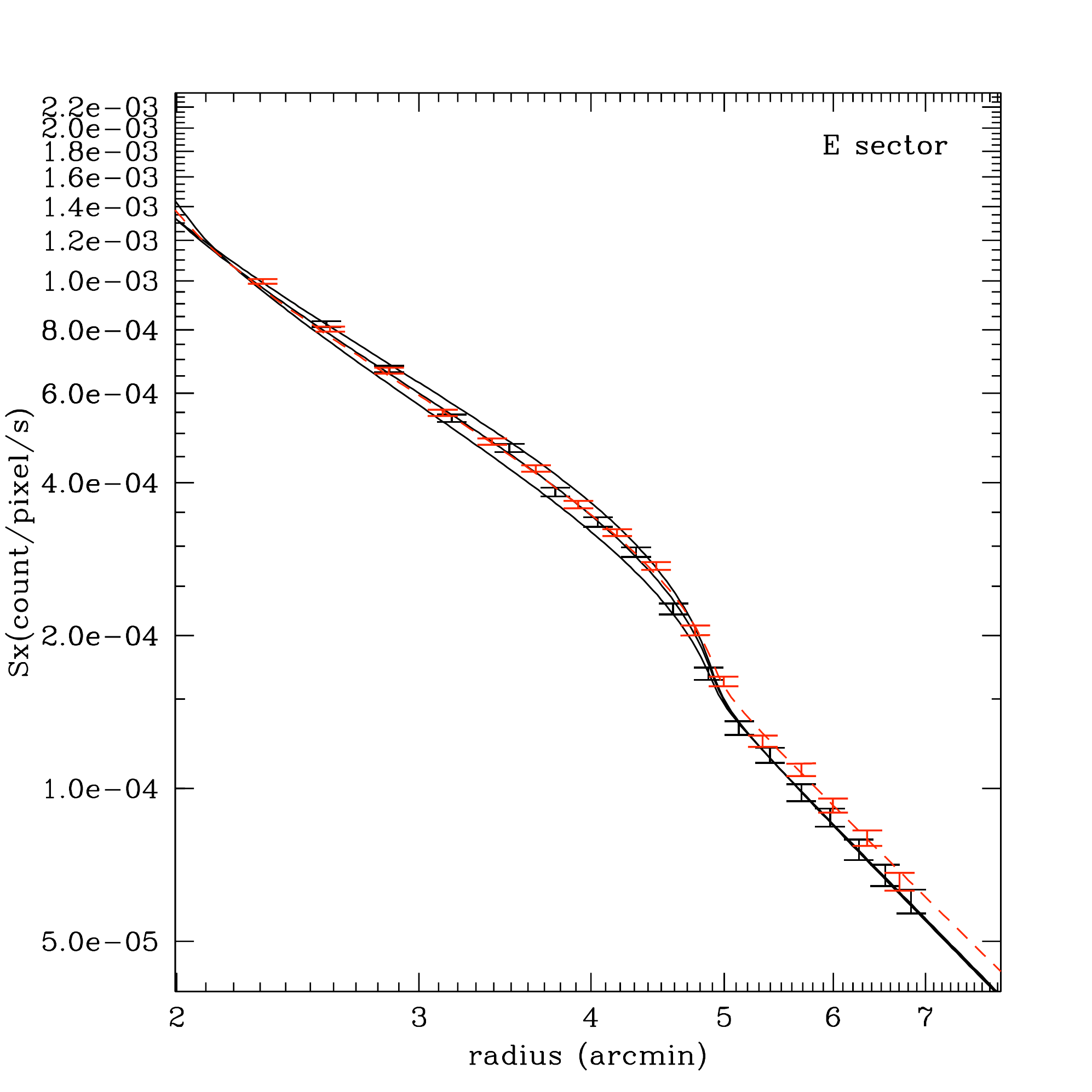}
\end{minipage}\\
\begin{minipage}{0.5\textwidth}
\includegraphics[width=\columnwidth]{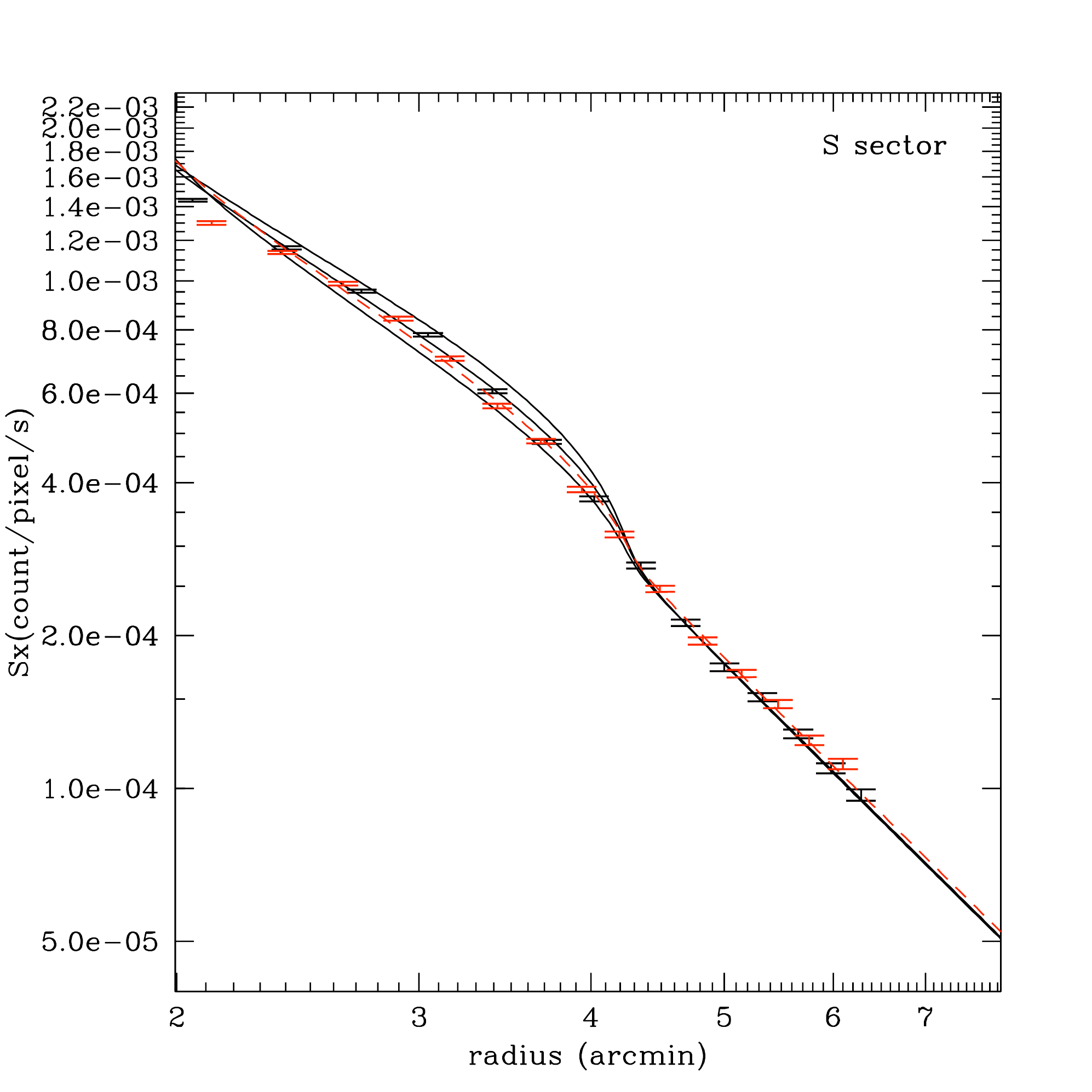}
\end{minipage}
\begin{minipage}{0.5\textwidth}
\includegraphics[width=\columnwidth]{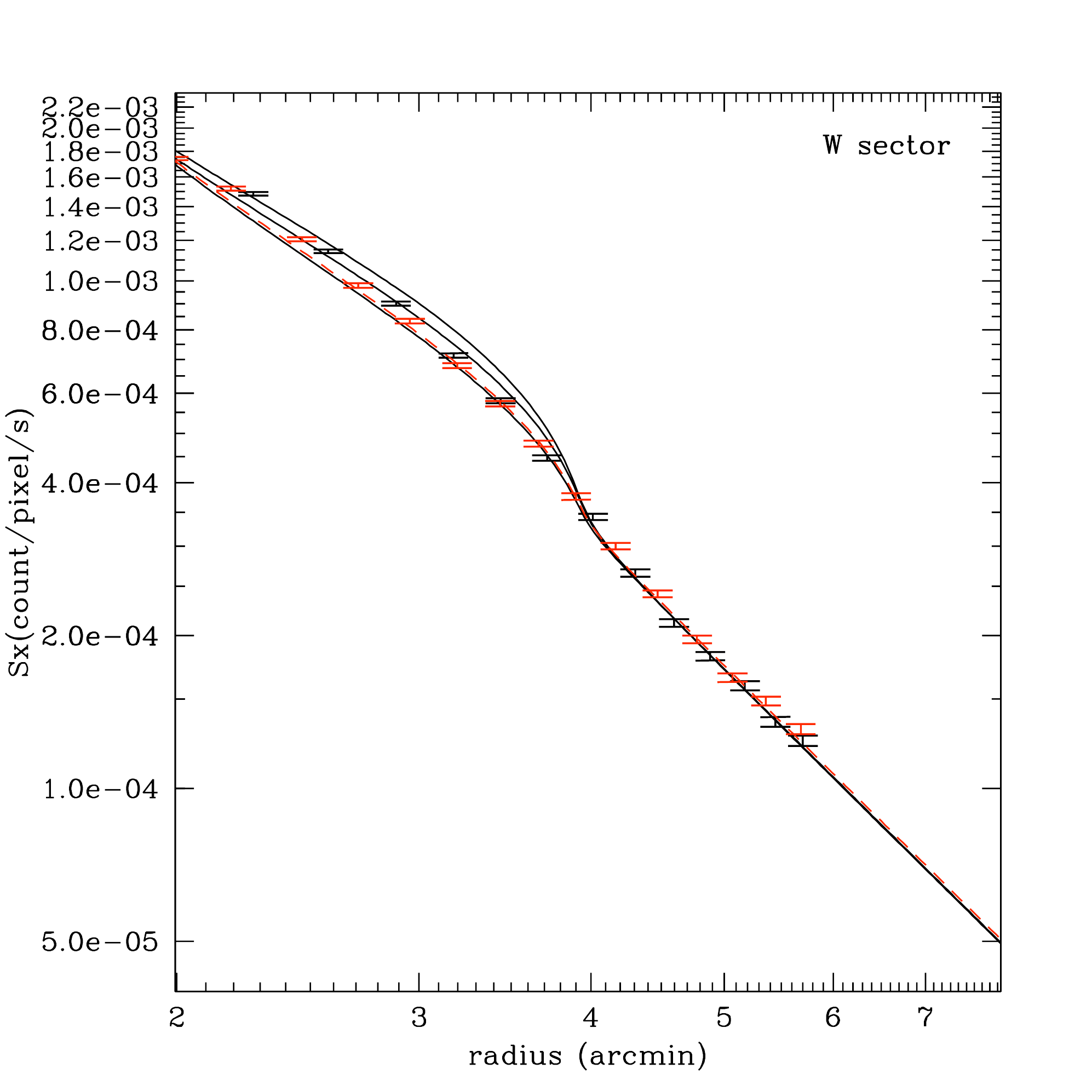}
\end{minipage}
\caption{Surface brightness profiles (from EPIC/pn) and fitted shock models (best-fit and bracketing). Data points from the pressure-map based annuli are indicated in black. The three lines show model surface brightness profiles for shocks with Mach numbers of 1.41, 1.30, 1.25 to the N; 1.35, 1.30, 1.25 to the E; 1.32, 1.27, 1.22 to the S; 1.35, 1.30, 1.25 to the W (increasing upward). 
Red points show the results from the Chandra-shape based annuli. For plot legibility, the bracketing models for the Chandra-shape based annuli are not shown. The model surface brightness profiles have Mach numbers of 1.33 to the N; 1.30 to the E; 1.23 to the S; 1.25 to the W.}
\label{fig:sxjumps}
\end{figure*}

To quantify the shock, we first use a spherically symmetric hydrodynamic model of a point explosion at the center of an initially isothermal, hydrostatic atmosphere \citep[as employed also by e.g.][]{Nulsen05}. We assume the initial gas density profile to be a power law, $\rho(r) \propto r^{-\eta}$, with $\eta$ adjusted to make the surface brightness profile of the undisturbed gas consistent with the observed surface brightness profile beyond the shock. The gravitational field, $g \propto 1/r$, and gas temperature are scaled to make the undisturbed atmosphere hydrostatic. 

The XMM-Newton 0.4 -- 7. keV response for a range of temperatures was tabulated using XSPEC, with detector response files appropriate for these observations and an absorbed mekal model with metal abundances set to 0.3 times solar (the results are insensitive to these parameters, including the preshock temperature). We scale the temperature profile from the model to obtain an unshocked gas temperature of 3.6 keV. The surface brightness profiles are then determined using the tabulated XMM response, the density and temperature model profiles, and assuming spherical symmetry to calculate the effects of projection along the line of sight. 
We additionally smooth the obtained profile by the XMM-Newton point-spread function (PSF), which we model by a King profile with a core radius of 3.5$^{\prime\prime}$ and a slope of 1.36 \citep{MarkevitchPSF}.

The shock weakens as the hydrodynamic model evolves and, since the initial conditions are self-similar, the flow can be scaled radially to place the model shock at the location of the observed shock. Surface brightness is finally scaled to match the observed profile in the unshocked region. In Fig. \ref{fig:sxjumps}, we show background-subtracted, exposure corrected EPIC/pn surface brightness profiles from elliptical annuli in four different sectors towards the N, E, S, and W from the core and the corresponding best-fit shock models, whose parameters are presented in Table \ref{tab:sxmach}. The error estimates in Table \ref{tab:sxmach} are based on the three (one best-fit and two bracketing) models plotted in Fig. \ref{fig:sxjumps} for each sector, respectively. 

We compare two different choices for the shape of these annuli: parallel to the ellipse which approximates the shock seen in the pressure map (Fig. \ref{fig:pmap}) and parallel to the four elliptical sectors which provide a more exact approximation to the shock seen in the Chandra surface brightness map (Fig. \ref{fig:chandrashock}).
A Mach 1.3 shock with an age of 130--230 Myr and an energy of 1.5--3$\times10^{61}$ ergs gives a reasonably good fit to the observed shock profile in each of the sectors using both choices of extraction region shape, in agreement with the values derived by \citet{Nulsen05}. Note that a Mach 1.3 model with a radius of approximately 5\arcmin\ (the average radius of the entire shock front) and age of 180 Myr also provides a reasonable fit to the integrated profile obtained by combining all the four sectors, apart from 2-3 data points immediately inside the shock front where, due to the non-spherical geometry, the jump appears smoother compared to the model prediction. The pre-shock electron density needed to calculate the shock energy was determined by fitting the data points outside 200 kpc in the deprojected profile of \citet{David01} with a power-law and extrapolating. 

\begin{table*}[!ht]
\begin{center}
\caption[]{Best-fit parameters for the shocked surface brightness model in the four sectors. Subscripts "p" and "c" denote the pressure- and Chandra-map based annuli shapes, respectively.}
\label{tab:sxmach}
\vspace{2mm}
\begin{tabular}{l|cccccc}
\hline
\hline
Sector & radius & $\eta$ & Mach & Shock energy & Shock age \\
 & (arcmin) & & number & ($10^{61}$ ergs) & (Myr)\\
\hline
Np	    & 6.0 & 2.9 & $1.30\pm0.08 $ & $2.3\pm1.0$ & $230\pm30$\\
Ep           & 5.0 & 2.8 & $1.34\pm0.06 $ & $2.6\pm0.6$ &$170\pm16$\\ 
Sp     	    & 4.3 & 2.7 & $1.30\pm0.06$ & $2.0\pm0.7$ & $154\pm20$\\
Wp          & 4.0 & 2.7 & $1.33\pm0.06$ & $2.3\pm0.7$ & $132\pm16$\\ 
\hline
Nc	    & 6.0 & 2.9 & $1.33\pm0.08 $ & $2.7\pm1.0$ & $215\pm30$\\
Ec         & 5.0 & 2.7 & $1.30\pm0.05 $ & $2.0\pm0.5$ &$180\pm16$\\ 
Sc    	    & 4.3 & 2.7 & $1.23\pm0.06$ & $1.4\pm0.6$ & $171\pm18$\\
Wc        & 4.0 & 2.7 & $1.25\pm0.06$ & $1.6\pm0.6$ & $151\pm18$\\ 
\hline
\end{tabular}
\end{center}
\end{table*}

Based on the best-fit radii in the different sectors, we can also obtain an independent estimate of the offset of the shock center with respect to the cluster center, which can be approximated as $\sqrt{(\frac{r_{\rm N}-r_{\rm S}}{2})^2+(\frac{r_{\rm E}-r_{\rm W}}{2})^2}\approx1^\prime$, within 10\% of the value estimated in Sect. \ref{sect:geom}.

There are several caveats in using this simple hydrodynamic model: the initial density profile is only well approximated as a power law locally, the shock front is clearly aspherical, and the outer radio lobes lie close behind the shock front in the N and S, so that they still have an influence in driving the shock, violating our assumption that the shock front is driven by a point explosion. The different shock ages required for different sectors (Table \ref{tab:sxmach}) also point out the need for a more complicated scenario. Despite these shortcomings, the models do provide reasonable fits to the surface brightness in the region of the shock front.

\section{The temperature jump associated with the shock}

\begin{figure}[!h]
\includegraphics[width=\columnwidth]{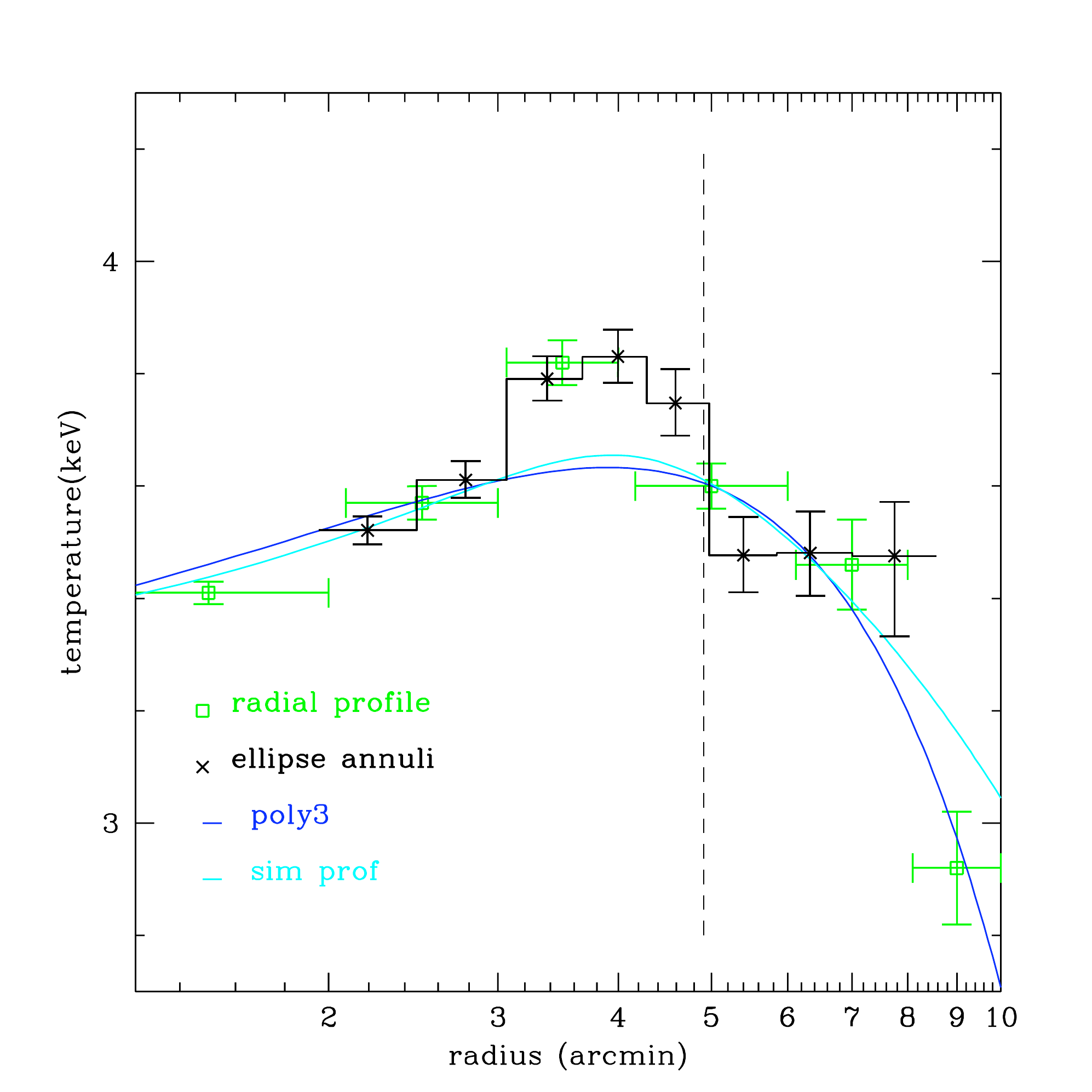}
\caption{Radial temperature profile from \citet{Simionescu_HydraAI}. Best-fit third degree polynomial over-plotted in blue, temperature profile assumed as initial condition for the simulations in Sect. \ref{sec:sim_parameters_cluster} in cyan. Temperature profile in elliptical annuli parallel to the shock seen in the pressure map as a function of average radius of the extraction region shown with black crosses. The average radius of the shock ellipse (Fig. \ref{fig:pmap}) is shown with a vertical dashed line.}
\label{fig:ktall}
\end{figure}

A hint of a temperature jump in Hydra A can already be seen in the radial profiles presented by \citet{Simionescu_HydraAI}. The data point in the 3--4\arcmin\ radial bin, behind the W and S shocks (located at 4 and 4.3\arcmin, Table \ref{tab:sxmach}), cannot be fit with a smooth function going through the other data points (Fig. \ref{fig:ktall}). The radial bins of \citet{Simionescu_HydraAI} were chosen independently of the shock in order to determine metallicity profiles; we further show in Fig. \ref{fig:ktall} the temperature profile obtained from elliptical annuli designed to follow the shock ellipse seen in the pressure map. These data points are plotted against the mean radius of the extraction regions. Note that each region contains emission from a broader range of radii than the circular annuli used for the radial profile. The average radius of the shock ellipse (Fig. \ref{fig:pmap}) is shown with a vertical dashed line. There is a good agreement with the radial profile and a temperature jump can be clearly seen behind the average radius of the shock ellipse. The temperature jump corresponds to an average Mach number M$=1.20\pm0.05$ (Table \ref{tab:tempmach}). The profile extracted from regions parallel to the Chandra-based shock shape is very similar and the temperature jump corresponds to M$=1.25^{+0.05}_{-0.10}$.

The temperature varies little between 2--8\arcmin, the region we are interested in for determining the shock properties. However, to eliminate any effects due to an underlying temperature profile in the cluster, we used a 3$^{\rm rd}$ degree polynomial fit to the radial temperature profile (excluding the point at 3.5\arcmin) to compare the temperature differences between the pre- and post-shock regions with expected temperature differences from this polynomial approximation of the cluster temperature profile. The polynomial fits the data points in the radial profile well (see Fig. \ref{fig:ktall}), but drops very rapidly at large radii (beyond the region of interest for determining the shock properties). For this reason, we adopt a different function for the radial dependence of the temperature for the 3D hydrodynamic simulations in Section \ref{sec:sim_parameters_cluster}. This function, over-plotted in cyan in Fig. \ref{fig:ktall}, agrees very well with the polynomial approximation in the region around the shock and shows a slower decrease at large radii, ensuring that the temperature does not become negative in the simulation box.

To describe the shock in more detail and to obtain temperature profiles using spectral extraction regions spanning a smaller range of radii, we also divided the data in four different sectors (N, E, S, and W) again using two sets of annuli for each sector: one parallel to the ellipse in the pressure map (following the white lines in Fig. \ref{fig:pmap}) and one parallel to the corresponding elliptical sector approximating the position of the Chandra surface brightness discontinuity (thin red lines in Fig. \ref{fig:chandrashock}). Towards the N and E we use wider annuli compared to the S and W, where the shock is closer to the cluster center providing a higher surface brightness around the shocked region. 

\begin{figure*}
\begin{minipage}{0.5\textwidth}
\includegraphics[width=\columnwidth]{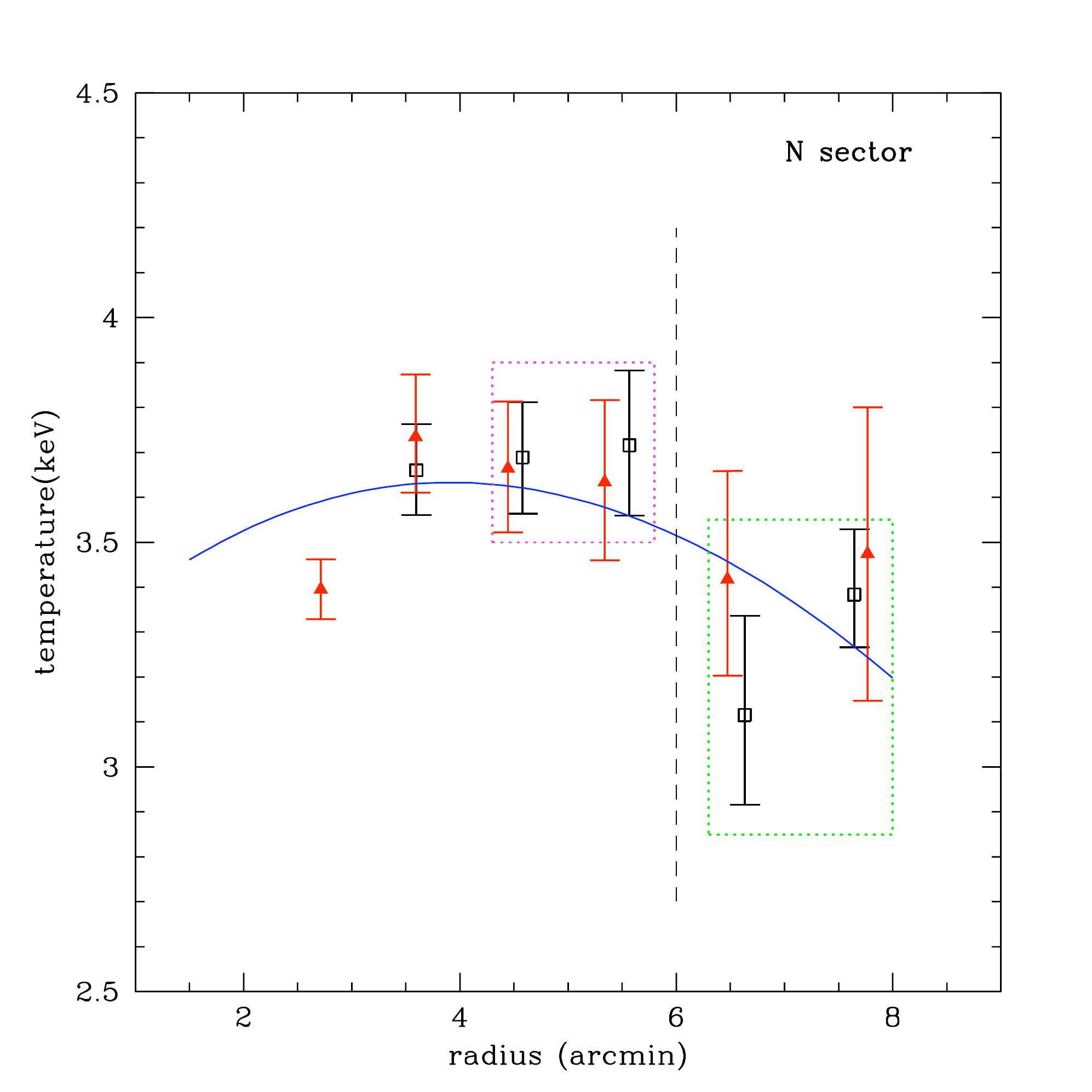}
\end{minipage}
\begin{minipage}{0.5\textwidth}
\includegraphics[width=\columnwidth]{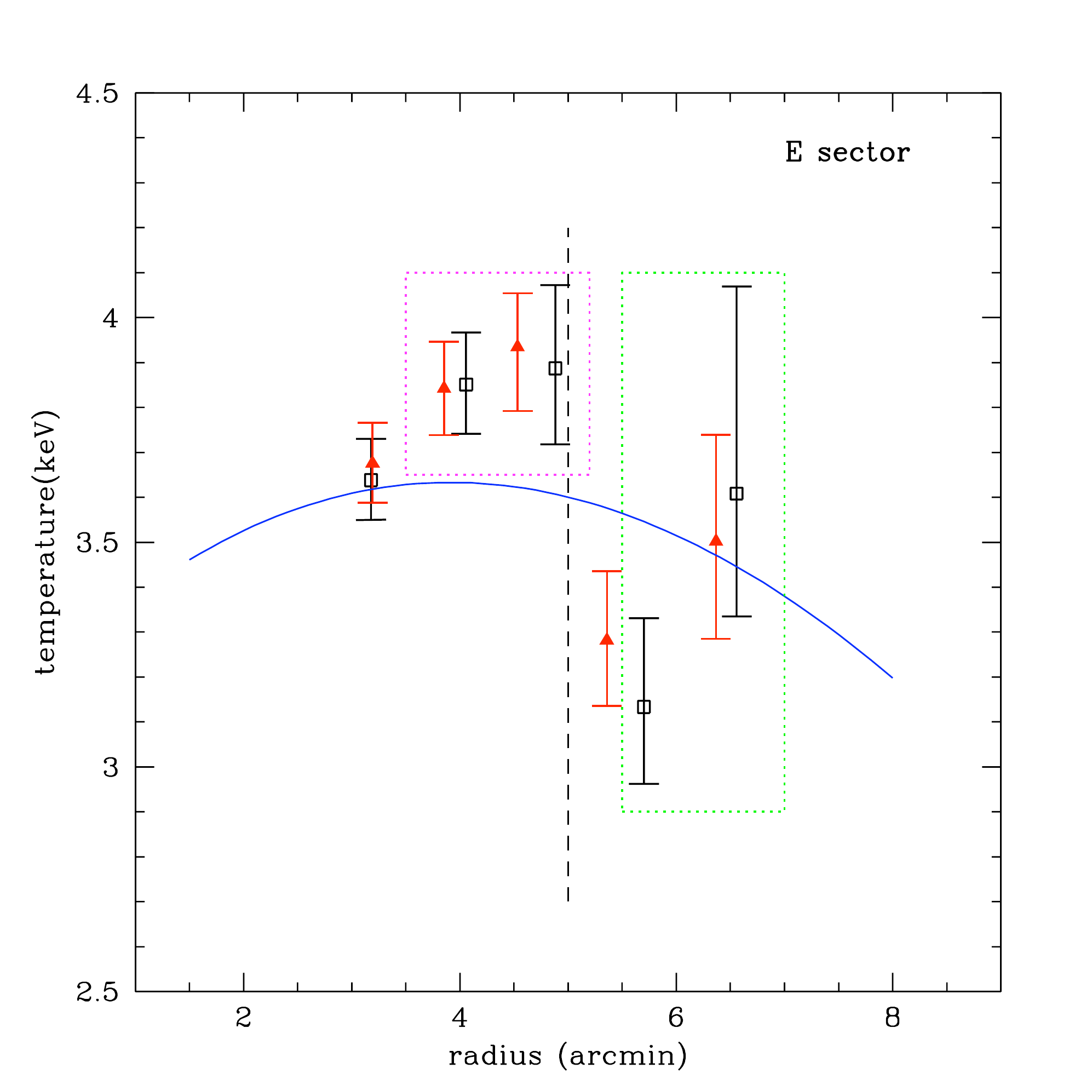}
\end{minipage}\\
\begin{minipage}{0.5\textwidth}
\includegraphics[width=\columnwidth]{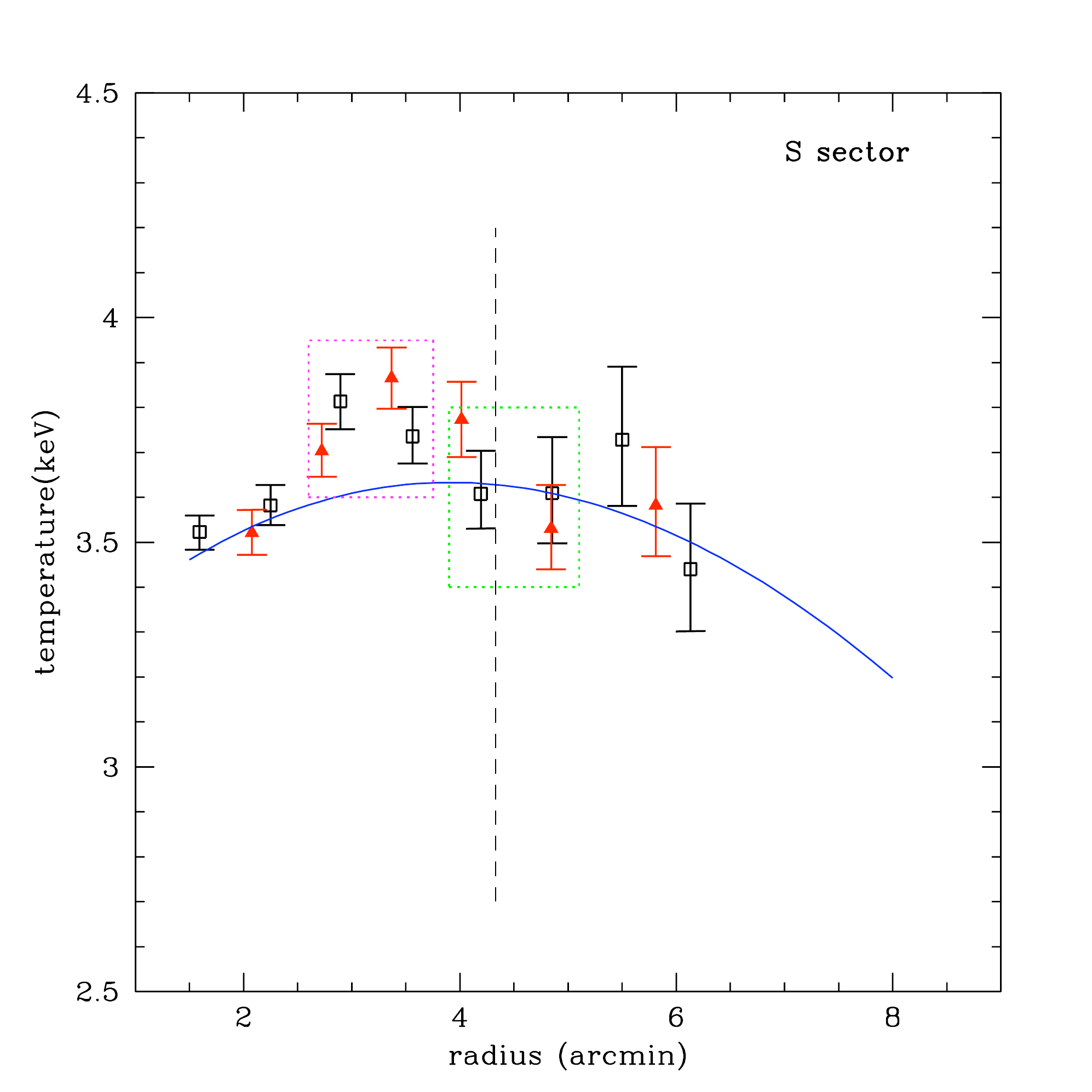}
\end{minipage}
\begin{minipage}{0.5\textwidth}
\includegraphics[width=\columnwidth]{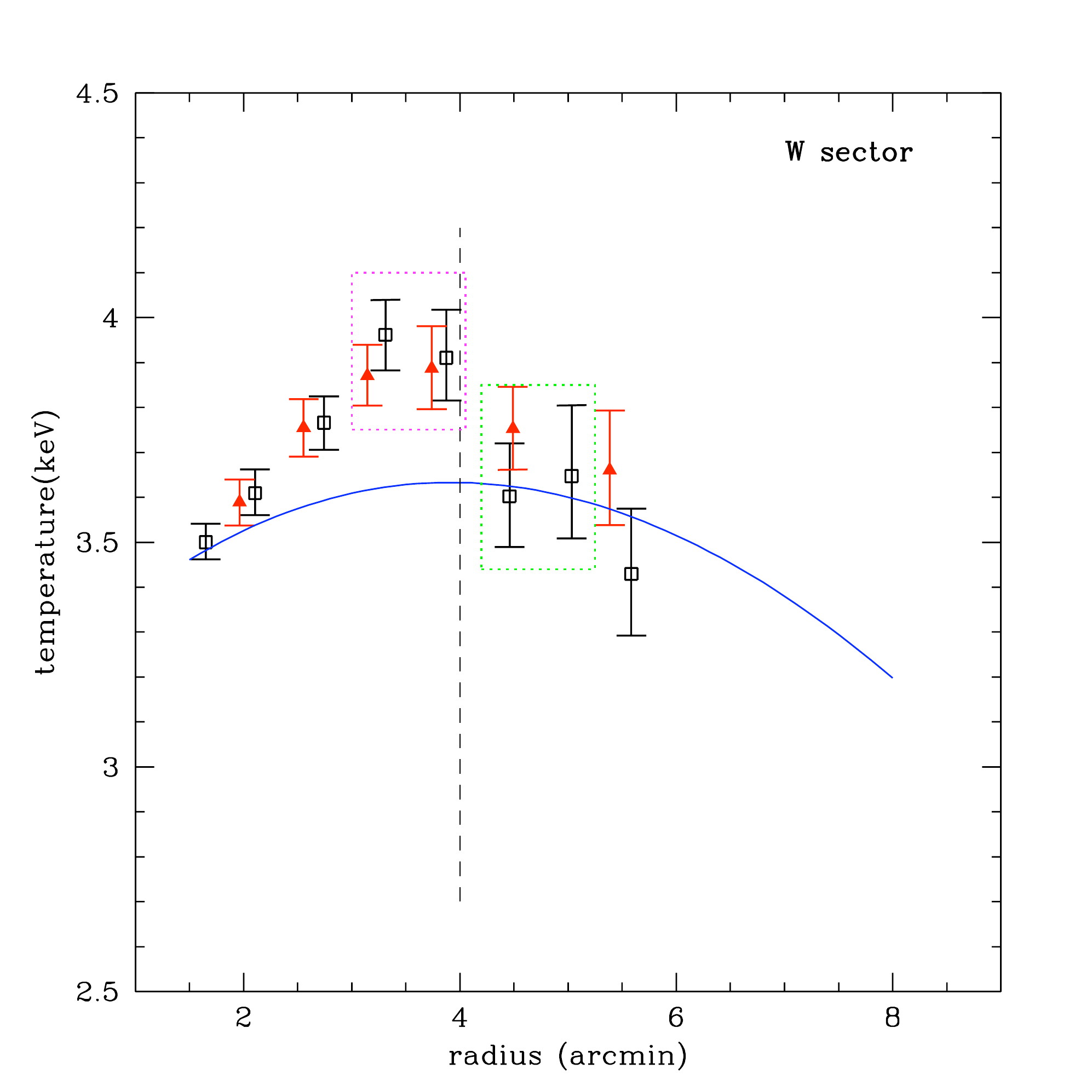}
\end{minipage}
\caption{Temperature profiles in 4 different sectors. Data points from the pressure-map based annuli are indicated with black squares, red triangles show the results from the Chandra-shape based annuli. The blue line represents a third-degree polynomial fit to the radial temperature profile. The dotted magenta and green rectangles indicate the pressure-map based annuli used to determine the temperature of the shocked gas and unshocked gas, respectively. The vertical dashed line indicates the position of the shock from the surface brightness fits (Table \ref{tab:sxmach}).}
\label{fig:tjumps}
\end{figure*}

The temperature profiles obtained for the two choices of extraction regions are in good agreement for all four sectors (see Fig. \ref{fig:tjumps}). 
The clearest indication of the presence of a temperature jump associated with the shock is seen in the W sector. Towards the south, a temperature jump is seen, but slightly further behind the surface brightness jump compared to the west (and the two different choices of extraction regions give different shifts). Towards the E, there is clear indication that the gas behind the shock is hotter than the expected average at that radius, but the temperature drop immediately beyond the shock raises the question of whether the used polynomial approximation is reliable for describing the undisturbed temperature distribution in this sector. Towards the N, there are only very weak ($<1\sigma$) deviations of the gas temperature in the shocked region from the radial average, and the errors in this sector are the largest given that it is furthest away from the cluster center and its opening angle is the smallest.

We grouped the two annuli with the largest jumps compared to the radial average to obtain a more accurate temperature of the shocked gas, k$T_{\rm sh}$, and the next two annuli at larger radii to obtain the temperature of the unshocked gas, k$T_{\rm unsh}$. For the elliptical shock shape approximation, the grouped annuli are marked in Fig. \ref{fig:tjumps}. For the Chandra-based regions, we always grouped the two annuli immediately before and after the vertical dashed line indicating the position of the shock from the surface brightness fits.
The results are shown in Table \ref{tab:tempmach}. We find a temperature jump at the position of the shock above that expected from a simple polynomial approximation of the cluster profile with a significance of typically 2$\sigma$ (between 0 and 2.9$\sigma$) in individual sectors (statistical only, neglecting the errors in determining the polynomial). The 1$\sigma$ error intervals for the temperature jumps obtained using the two different region choices overlap in each of the 4 sectors.
Combining the significances in the four sectors, the total significance of the temperature jump associated with the shock is 4.3$\sigma$ for the pressure-based regions and 4.0$\sigma$ for the Chandra-based regions. This is higher than the significance obtained fitting the entire annuli without dividing into sectors. A possible explanation is that for large, asymmetric regions spanning a wide range of radii, the average temperature is not so well represented by evaluating the polynomial fit to the radial temperature profile at the average radius of those regions, which may overestimate the expected temperature differences in the absence of a shock ($\Delta T_{\rm exp}$ in Table \ref{tab:tempmach}) and thus underestimate the temperature jump. This effect is less important if the extraction regions are smaller (e.g. the sectors) and cover a smaller range in radius. 

\begin{figure}
\includegraphics[width=\columnwidth]{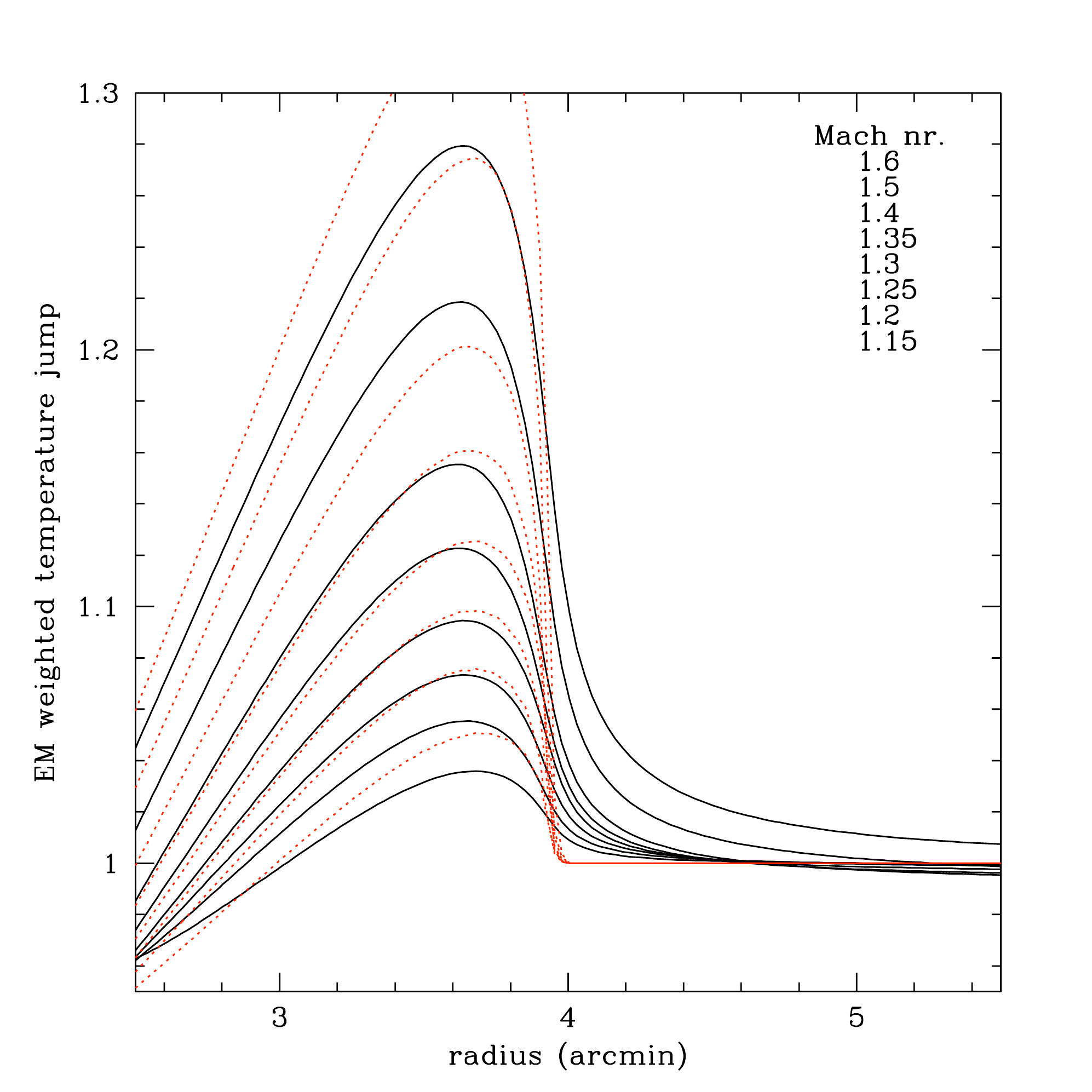}
\caption{Expected emission-measure weighted, projected temperature profile across the shock relative to the undisturbed temperature using the 1D isothermal model. The red dotted line includes only projection effects, the black solid line additionally takes into account the smoothing due to the XMM-PSF. Model Mach numbers range from 1.15 to 1.6 from bottom to top, as example $\eta$ was chosen to be 2.7.}
\label{fig:emwt}
\end{figure}

To calculate the Mach number from the observed temperature jumps, we resort again to the 1D shock model. We use the 1D temperature and density model profiles and the assumption of spherical symmetry to determine an emission-measure weighted temperature profile which takes into account the effects of projection along the line of sight. We then apply the correction due to the XMM-PSF. By plotting the temperature profiles obtained in this way for models with several Mach numbers (Fig. \ref{fig:emwt}), we can identify which Mach number corresponds to the temperature jump (in percent) which we observe. These are also reported in Table \ref{tab:tempmach}. The relative temperature jumps are calculated with respect to the expected temperature in the shocked region in the absence of a shock, which is determined from the polynomial approximation to the radial temperature profile.

\begin{table*}[tb]
\begin{center}
\caption[]{Best-fit post- and pre-shock temperatures, temperature jump $\Delta T$, expected difference from the radial dependence of the cluster temperature $\Delta T_{exp}$, relative temperature jump, and corresponding Mach number in the four sectors. Subscripts "p" and "c" denote the pressure- and Chandra-map based annuli shapes, respectively.}
\label{tab:tempmach}
\vspace{2mm}
\begin{tabular}{l|cccccc}
\hline
\hline
Sector & k$T_{\rm sh}$ & k$T_{\rm unsh}$ & $\Delta T$ & $\Delta T_{\rm exp}$ & $\frac{\Delta T-\Delta T_{\rm exp} } {T_{\rm sh,exp} }$ & Mach \\
\hline
All p  & $3.80^{+0.03}_{-0.03}$ & $3.56^{+0.05}_{-0.05}$ & $0.24\pm0.06$ & 0.10 & $3.9\pm1.7$\% & $1.20^{+0.05}_{-0.05}$  \\ 
\hline
Np	    & $3.73^{+0.09}_{-0.10}$ & $3.22^{+0.18}_{-0.17}$ & $0.51\pm0.20$ & 0.24 & $7.5\pm5.6$\% & $1.35^{+0.10}_{-0.20}$ \\
Ep           & $3.86^{+0.10}_{-0.09}$ & $3.33^{+0.17}_{-0.15}$ & $0.53\pm0.19$ & 0.12 & $11.3\pm5.2$\% & $1.40^{+0.05}_{-0.15}$  \\ 
Sp     	    & $3.79^{+0.04}_{-0.05}$ & $3.62^{+0.07}_{-0.07}$ & $0.17\pm0.08$ & 0.00 & $4.7\pm2.2$\% & $1.25^{+0.05}_{-0.10}$ \\
Wp          & $3.95^{+0.07}_{-0.06}$ & $3.63^{+0.10}_{-0.09}$ & $0.32\pm0.11$ & 0.02 & $8.3\pm3.0$\% & $1.30^{+0.10}_{-0.05}$  \\ 
\hline
All c  & $3.85^{+0.03}_{-0.03}$ & $3.60^{+0.05}_{-0.05}$ & $0.25\pm0.06$ & 0.09 & $4.5\pm1.7$\% & $1.25^{+0.05}_{-0.10}$  \\ 
\hline
Nc	    & $3.71^{+0.11}_{-0.11}$ & $3.44^{+0.18}_{-0.17}$ & $0.51\pm0.20$ & 0.25 & $<6.4$\% (1$\sigma$)& $<1.35$ \\
Ec         & $3.87^{+0.08}_{-0.07}$ & $3.35^{+0.12}_{-0.12}$ & $0.53\pm0.19$ & 0.11 & $11.3\pm3.9$\% & $1.40^{+0.10}_{-0.10}$  \\ 
Sc    	    & $3.84^{+0.05}_{-0.05}$ & $3.57^{+0.08}_{-0.07}$ & $0.17\pm0.08$ & 0.06 & $5.8\pm2.5$\% & $1.25^{+0.10}_{-0.05}$ \\
Wc        & $3.90^{+0.06}_{-0.06}$ & $3.73^{+0.08}_{-0.08}$ & $0.32\pm0.11$ & 0.03 & $3.9\pm2.6$\% & $1.20^{+0.10}_{-0.10}$  \\ 
\hline
\end{tabular}
\end{center}
\end{table*}

\section{Relative motions in the cluster and the shape of the shock front}

In Sect. \ref{sect:geom}, we showed that the shock front is elliptical and that its center is offset from the cluster center by $\sim$70 kpc. However, in Sect. \ref{sect:mach}, we conclude that the Mach number is roughly the same for all the four different sectors, although the distance between the shock front and the cluster center differs. This is puzzling because, if the shock originates at the cluster center - which currently coincides roughly with the position of the AGN - then the shock front must have been moving faster on average towards the N, where it is now further away from the center, than towards the S or W. If the shock was moving as much as 1.5 times faster on average in one direction than in another (the ratio of the shock radii in the N and W, Table \ref{tab:sxmach}), one would expect to see some differences also in the current distribution of Mach numbers along the shock front: scaling up the Mach number in the W by a factor 1.5 means the expected Mach number in the N should be around 1.8, which is excluded by the data. The discrepancy between the model shock ages in different sectors also hints at the same problem, namely that the observed shape and properties of the shock cannot be simply explained by a point explosion.

Not only the shock front but also the radio lobes show a N--S asymmetry. While the shock radius to the N is 1.5 times bigger than to the S, the N radio lobe also extends much further out than the S lobe, which rises and then seems to bend back towards the cluster center. This led us to investigate whether relative motions between the AGN and the ICM could explain both the disturbance of the southern radio lobe and the elliptical, offset shape of the shock front.

To this end, we performed 3D hydrodynamical simulations of a symmetrical pair of back-to-back jets that originate from the cluster center. These jets interact with the surrounding ICM in which we triggered large-scale motions. 
\begin{figure}
\centering\resizebox{\hsize}{!}{\includegraphics[width=0.32\textwidth]{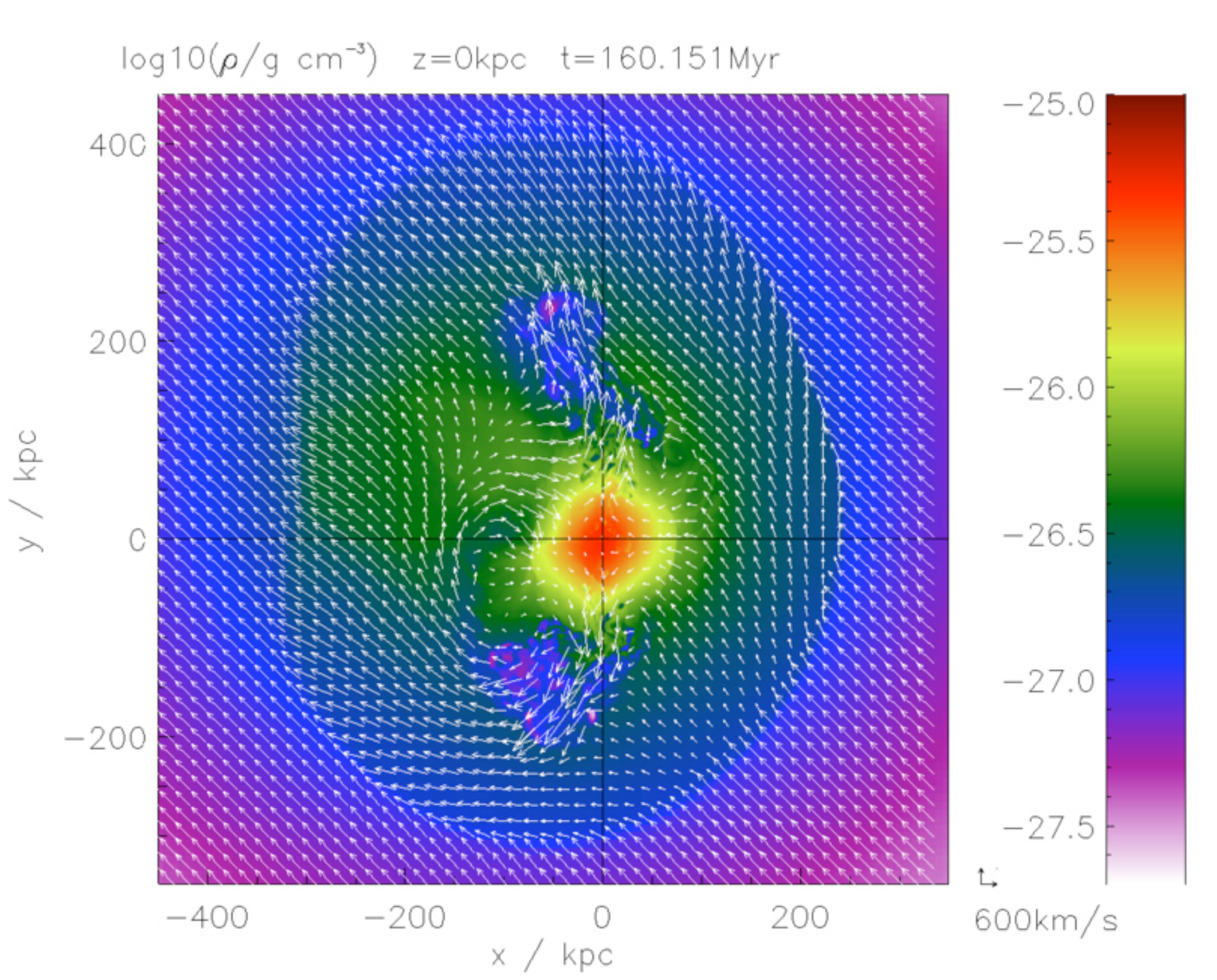}}
\centering\resizebox{\hsize}{!}{\includegraphics[width=0.32\textwidth]{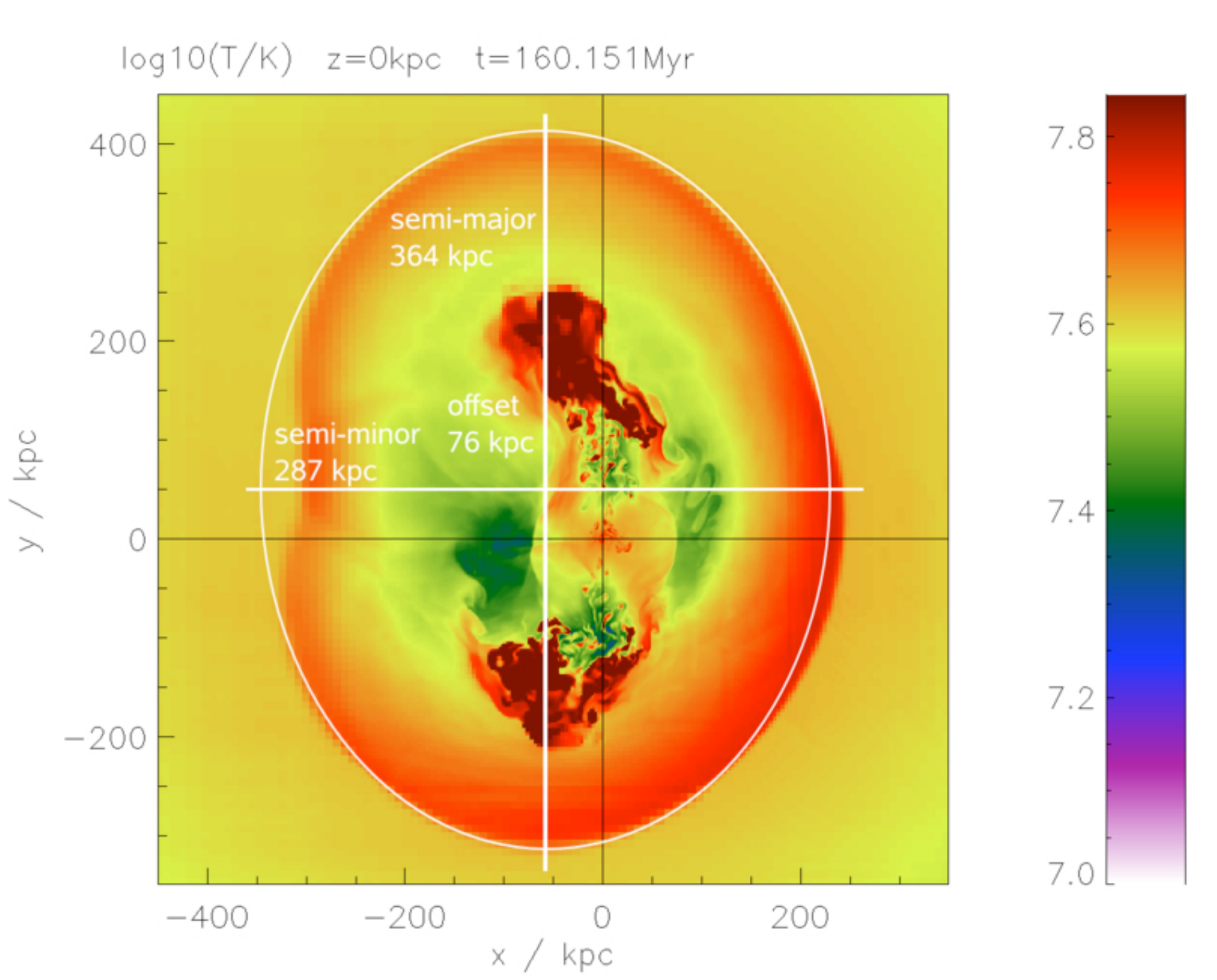}}
\centering\resizebox{\hsize}{!}{\includegraphics[width=0.32\textwidth]{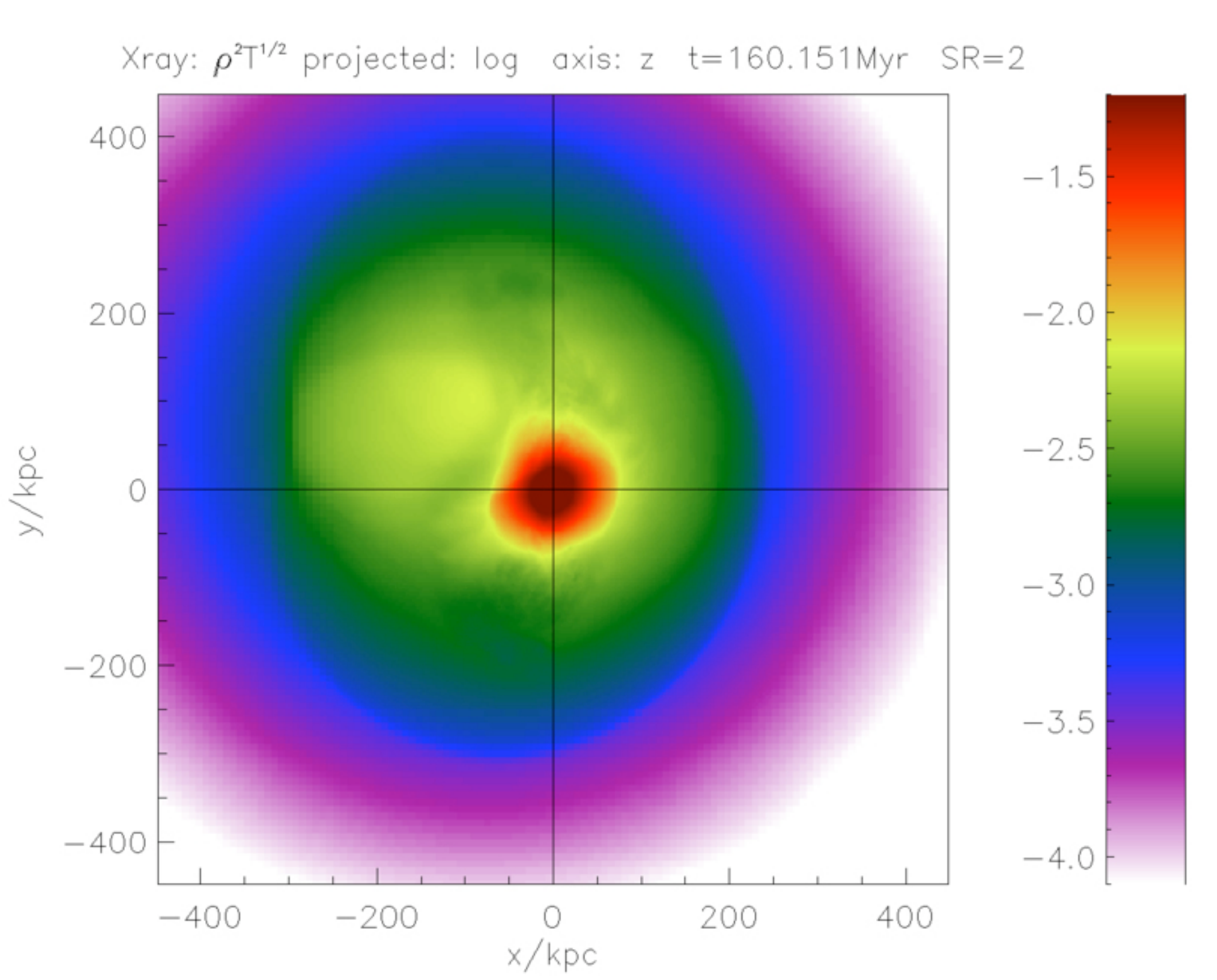}}
\caption{Shock front and bubbles at 160 Myr. {\it Top:} slice through the
computational domain, showing density (color-coded) and
velocity field (arrows). {\it Middle:} temperature (color-coded), with fitted shock
ellipse (white line) and labelled axes and offset. {\it Bottom:} Projected X-ray emission from simulated cluster.}
\label{fig:slices}
\end{figure}

\subsection{Model Cluster} \label{sec:sim_parameters_cluster}
We assume the gas density and temperature distribution in the cluster to be spherically symmetric. The initial radial density profile for the ICM is modelled by a double $\beta$-profile
(see \citealt{Wise07} and \citealt{David01}). The parameters are given in Table~\ref{tab:ICM}.
%
\begin{table}
\caption{ICM parameters used for 3D simulation.}
\label{tab:ICM}
\centering\begin{tabular}{lcc}
\hline
                    & component 1 & component 2\\
\hline
$R\ICM$ ($\Kpc$)        &  $27.7$  & $235.6$\\
$\rho\ICM{}_0$ ($\gccm$) &  $8\cdot 10^{-26}$ & $3.33\cdot 10^{-27}$\\
$\beta$        &  $0.686$ &  $0.907$ \\
$m_1$ $(\K/\Kpc)$   & \multicolumn{2}{c}{25,000} \\
$n_1$ ($\K$)          & \multicolumn{2}{c}{$3.6\times 10^7$}\\
$m_2$ $(\K/\Kpc)$    & \multicolumn{2}{c}{$-21,000$}\\
$r\Break$ ($\Kpc$)      & \multicolumn{2}{c}{250}\\
$a\Break$ ($\Kpc$)      & \multicolumn{2}{c}{100}\\
\hline
\end{tabular}
\end{table}
%
As a temperature profile, $T(r)$, we use
\begin{eqnarray}
T(r)&=&T_1(r)C(r) + T_2(r)[1-C(r)], \textrm{where}\\
T_1(r)&=&m_1 \cdot r + n_1 \nonumber\\
T_2(r)&=&m_2 \cdot r + n_2 \nonumber\\
n_2&=&(m_1-m_2)r\Break + n_1 \Leftrightarrow T_1(r\Break)=T_2(r\Break) \nonumber\\
C(r)&=&1-\frac{1}{\pi}\left(\arctan\frac{r-r\Break}{a\Break}+\frac{\pi}{2}\right)\nonumber,\\
\end{eqnarray}
which connects two linear functions smoothly. The parameters (see Table~\ref{tab:ICM}) are chosen to fit the data of \citet{Simionescu_HydraAI}. The temperature is declining both inwards and outwards from a cluster-centric radius of 250 kpc.
For the initial cluster we assume hydrostatic equilibrium, so that the given ICM density and temperature profiles determine the pressure distribution, and thus the gravitational potential in the cluster. 

\subsection{Model Jet} \label{sec:sim_parameters_jet}

The jet pair is modelled by defining two jet ``nozzles'' composed of all grid cells immediately below and above the $x$-$z$-plane, whose
distance to the $y$-axis falls below the jet nozzle radius, $r\Jet$. For the grid cells immediately above the $x$-$z$-plane, the fluxes towards the
$+y$-direction are set to a strong outflow with density $\rho\Jet$, pressure $p\Jet$ and velocity $v\Jet$. The grid cells immediately below the
$x$-$z$-plane are treated accordingly to produce the flow in the $-y$-direction.
We use the parameters listed in Table~\ref{tab:jet}.
%
\begin{table}
\caption{Jet parameters used for 3D simulation.}
\label{tab:jet}
\centering\begin{tabular}{ll}
\hline
radius $r\Jet$      &$6\Kpc$ \\
power $P\Jet$       &  $4.75\times 10^{46} $  erg$/s$  \\
lifetime $\tau\Jet$ & $10\Myr$ \\
total energy input      & $3\times 10^{61}$ erg\\
velocity $v\Jet$    & $20\,000\Kms$ \\
Mach number $M\Jet$ & $5.9$ \\
\hline
\end{tabular}
\end{table}
%
The chosen Mach number brings the jet gas close to pressure equilibrium with
the ICM in the cluster center, so that during the active phase the jet is mainly momentum driven.

\subsection{Generation of bulk motion}
The ICM in galaxy clusters is known to be in motion, the velocity field being both inherited from cluster formation and more recently achieved by current gas accretion flows. \citet{Heinz06} showed that these motions can have an important impact on the buoyant rise of radio bubbles. Although such flows are not necessarily coherent over scales of several 100 kpc, in Hydra A the offset of the shock ellipse and morphology of the radio lobes suggest a large-scale predominantly coherent flow. As a most simple model for such a scenario, we initialize a smooth velocity field in the ICM of our model cluster, namely a potential flow around a sphere of a 100 kpc radius, centered on the cluster center. This choice has the advantage that the ICM in the cluster center is only mildly affected by the bulk motion, but the shock front and the bubbles spend enough time in the flow region to be affected. Furthermore, the direction and amplitude of the bulk flow are clearly defined by requiring that when the shock reaches the observed size and age, its center is offset from the cluster center by the observed distance of 70 kpc towards the NE (Sect. \ref{sect:geom}). We use a flow velocity of $670 \Kms$ towards $(-1,1,0)$ based on the simple analytical argument that exposing the shock to a bulk flow of $670\Kms$ for about $100\Myr$ (the approximate estimated age minus the time it takes for the shock to reach the flow region), it would be carried along with the flow by $670\Kms \times100\Myr\approx70\Kpc$. This assumes that the bulk flow is in the plane of the sky; if the actual gas motions make an angle $\theta$ with respect to this plane, one would need a higher velocity of 670/cos$\theta \Kms$.

\subsection{Code}
We use the Eulerian adaptive-mesh refinement code FLASH (version 2.5) with radiative cooling \citep{sutherland93} and static gravity. The cluster potential is defined by the parameters described in Sect.~\ref{sec:sim_parameters_cluster}. The jet module is similar to the one in \citet{brueggen07a}.
 
The total grid size is $(-1\Mpc,1\Mpc)^3$. The adaptive mesh refinement allows
us to use a resolution of $0.5\Kpc$ in the cluster center, which ensures that the jet nozzle is resolved sufficiently. In order to limit computational
requirements, we restrict the refinement with increasing distance to the cluster center. The best resolution allowed outside 16, 100, and 200 kpc is 1, 4, and 8 kpc, respectively. We note that the lowest achieved resolution is on the order of the XMM-PSF. 

\subsection{Simulation results}
%
\begin{figure*}
\includegraphics[width=0.49\textwidth, bb=50 50 554 301]{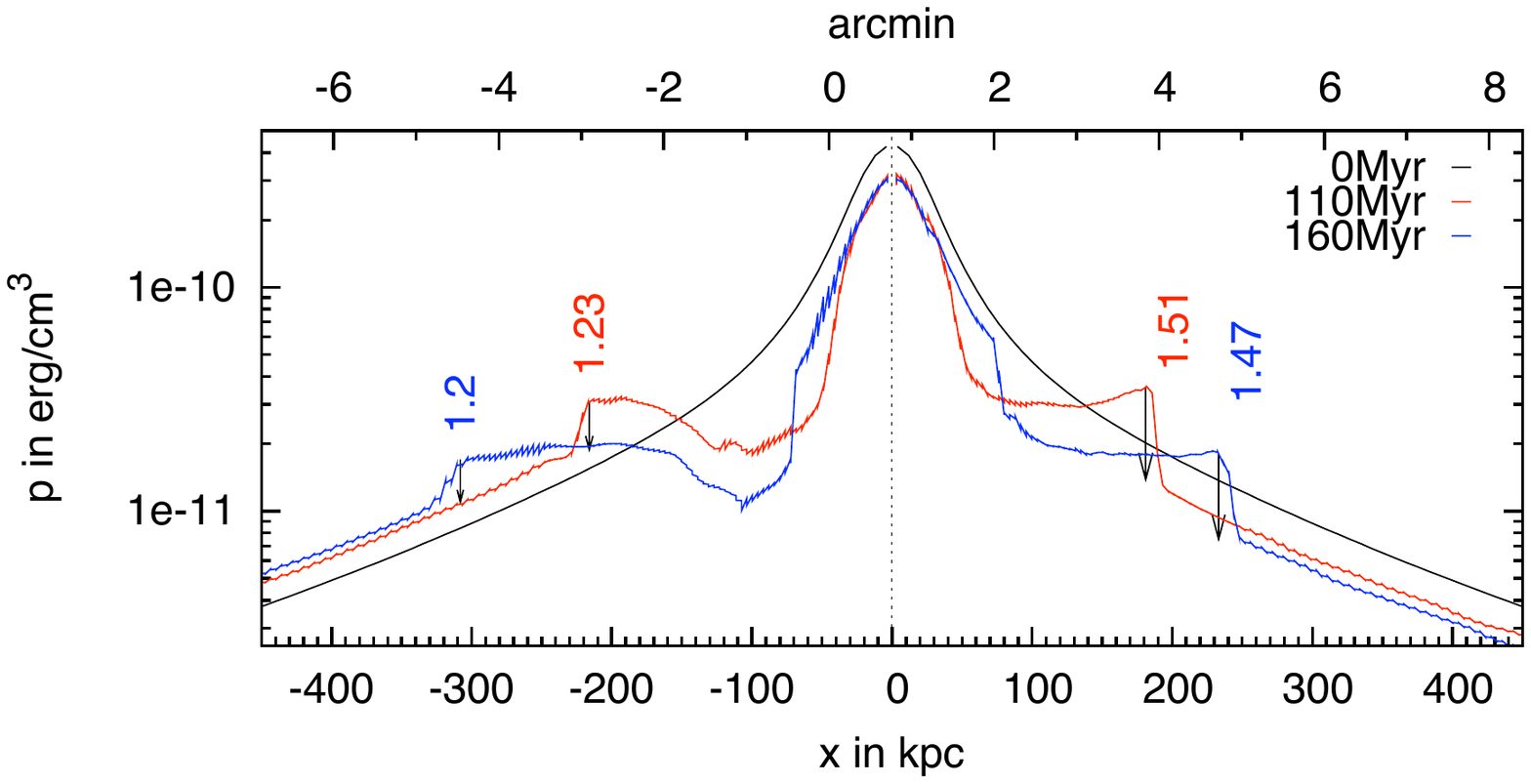}
\includegraphics[width=0.49\textwidth, bb=50 50 554 301]{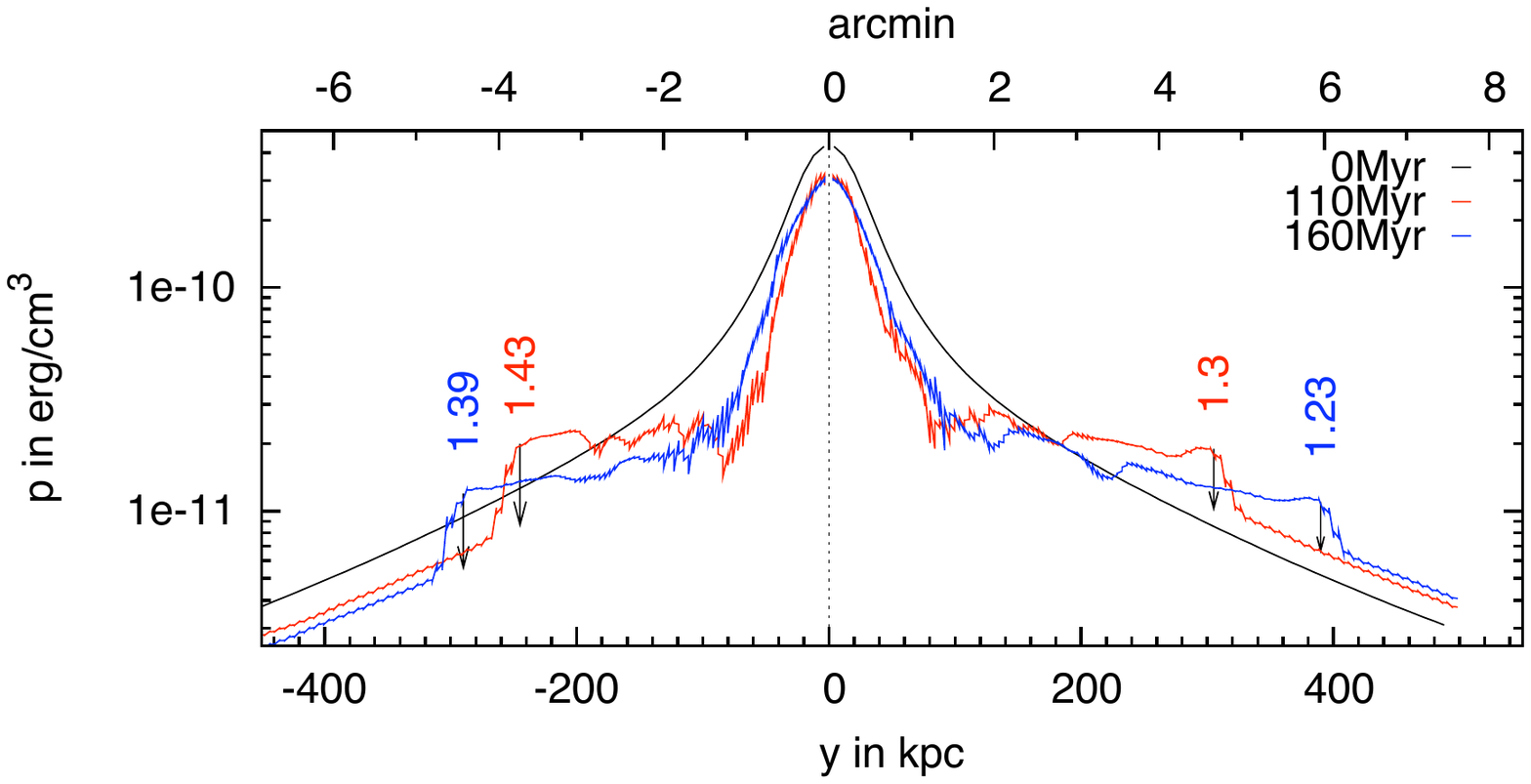}
\caption{Evolution of pressure profiles along the grid's $x$- and
$y$-axis through the center of the cluster. The shock is marked by the black arrows, numbers of corresponding
color indicate the Mach numbers of the shock as derived from the pressure jump.}
\label{fig:prof_evol}
\end{figure*}
The evolution of the jet-induced structures proceeds in two stages: the active jet phase and the subsequent passive phase (compare also \citealt{reynolds01}).  During the
active phase, the jets evacuate two channels along the $y$-axis. During this interval, the jet-induced structure still resides inside the inner 100 kpc
and thus does not feel the bulk flow.  When the jets are switched off, the evolution proceeds passively.  At that moment, the jet cocoons are still
overpressured and continue to expand and drive a shock into the surrounding
ICM. The expanding cocoons form two bubbles which rise buoyantly. While doing so, in their wakes, they drag up cooler gas from the cluster
center. Eventually, the shock detaches from the bubbles, while the shock is still expanding and the bubbles are still rising. At $160\Myr$, the shock has reached the size observed in Hydra A.

About 50 Myr after the jet was started, the shock enters the region of the
bulk flow, which causes different shock propagation speeds with respect to the
cluster center for different directions. Thus, after 160 Myr, the shock ellipse
is offset approximately towards $(-1,1,0)$ from the cluster center by 70 kpc,
as shown in Fig.~\ref{fig:slices}. This figure shows a density and a
temperature slice through the grid at $160\Myr$ as well as the corresponding simulated X-ray map obtained by integrating $\rho^2\sqrt{T}$ along the z direction.

In Fig.~\ref{fig:prof_evol}, we show pressure profiles along the grid's $x$-and $y$-axes. The discontinuities due to the shock can be clearly
seen. From the ratio of the pre- and post-shock values we calculated a Mach
number for each profile, resulting in an average Mach number of 1.3 at $160\Myr$. The Mach numbers in different directions range between about 1.2 and 1.5.

\section{Discussion}

The simulation successfully provides one possible explanation for the problem posed at the beginning of the previous section, namely why the Mach number is roughly the same in different sectors although the distance between the shock front and the cluster center differs. The simulated shock has Mach numbers and radii which agree with the observations in all sectors, and requires a consistent shock age in all directions, which a simple point explosion did not. 

\subsection{Comparison to observation}
The simulation successfully reproduces:
\begin{itemize} 
\item the size and shape (ellipticity) of the observed shock. The ellipse we fitted to the simulated shock (see
Fig.~\ref{fig:slices}) has a semi-major axis of $\approx$365 kpc and a semi-minor axis
of $\approx$290 kpc. 
\item the observed overall shock Mach number of around $M\sim1.3$.
\item the observed offset of the shock ellipse, about $70\Kpc$. 

\item a total energy input and average shock age of $3\times10^{61}$ ergs and 160 Myr, respectively. These are within the range of the values derived from the 1D model. 
\end{itemize}

\begin{figure}
\centering\resizebox{\hsize}{!}{\includegraphics[width=0.32\textwidth]{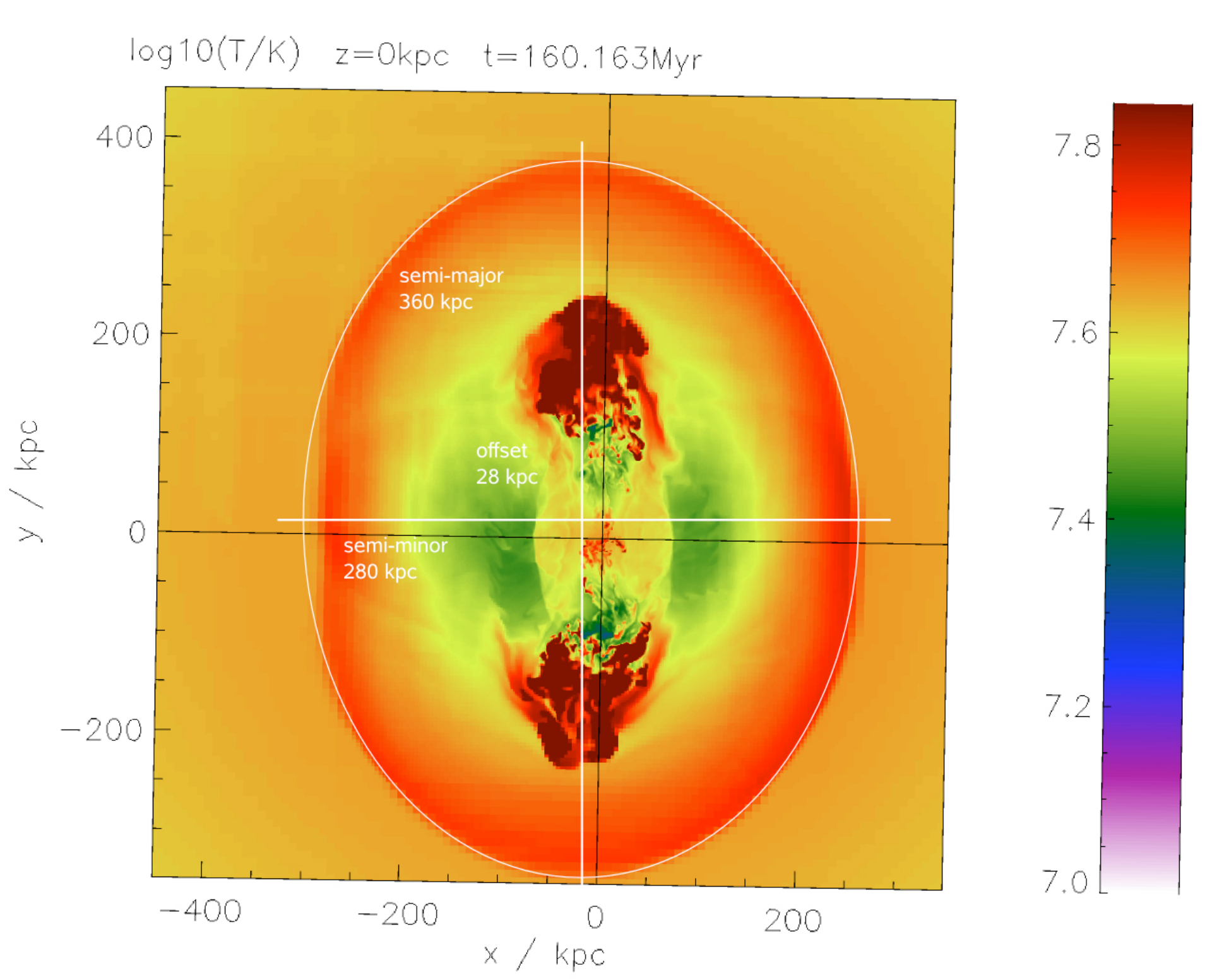}}
\centering\resizebox{\hsize}{!}{\includegraphics[width=0.32\textwidth]{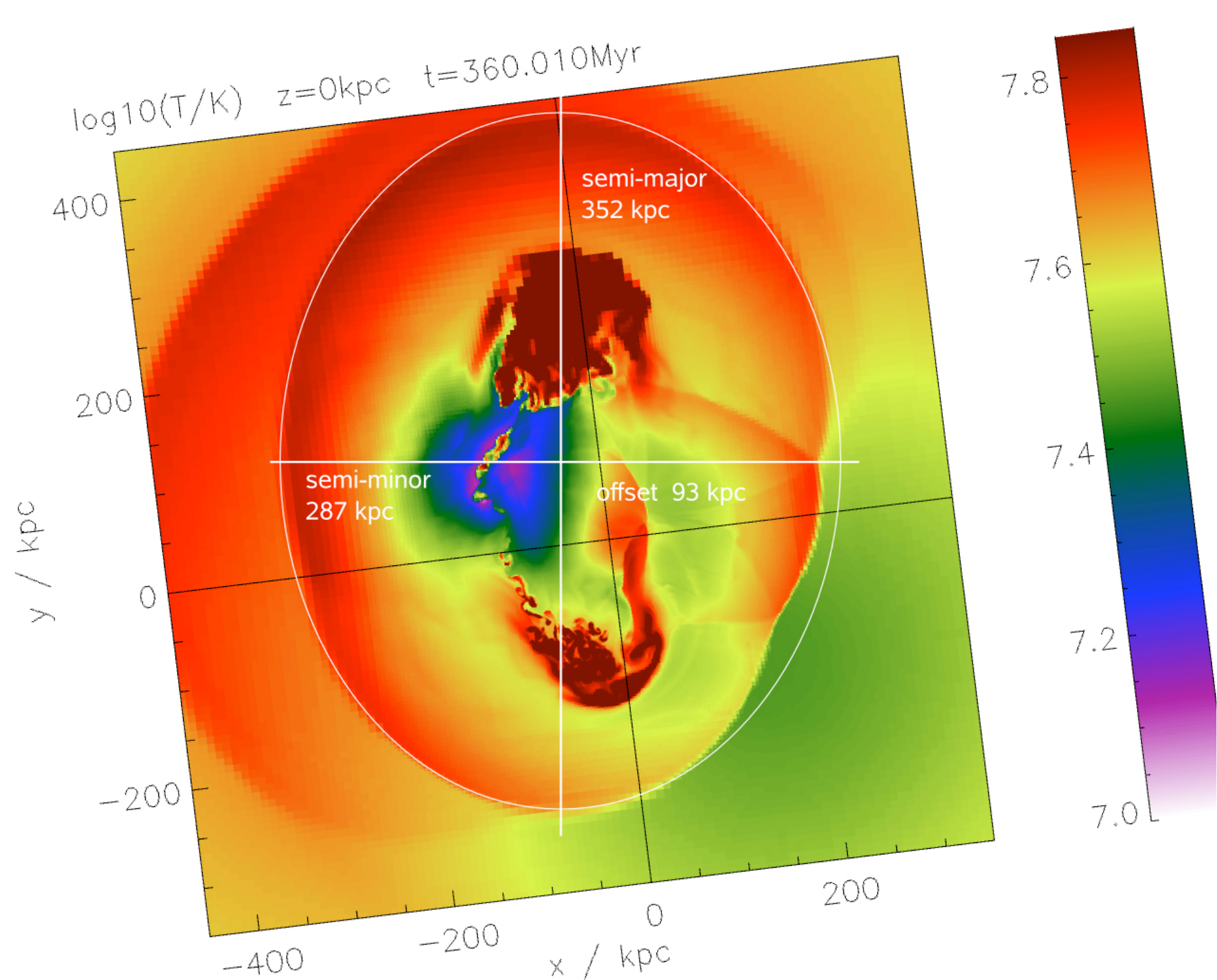}}
\caption{Temperature slices taken at $160\Myr$ after the jet was launched for
two alternative simulations. {\it Top:} the bulk flow velocity was reduced to
$260\Kms$. {\it Bottom:} The ICM bulk flow was generated by ICM sloshing.}
\label{fig:slices_altern}
\end{figure}

We see, therefore, an encouraging agreement with the observations. Several drawbacks, on the other hand, are listed below:

\begin{itemize} 
\item a bulk flow in the ICM leads to different Mach numbers at opposite sides
of the shock: Where the shock has to move against the ICM flow, its effective velocity with respect to the ICM is higher and thus the shock is stronger (in this simulation, in the $-y$- and $+x$-direction, respectively S and W). The observational data seems to indicate the opposite, namely a Mach number which is stronger in the E than in the W. However, with the current error bars, the observational results are still consistent with the simulation.

\item the strong bulk flow affects the shape of the simulated southern radio lobe but does not bend it nearly as much as the observed lobe.

\item  In the simulation, the shock is detached from the northern bubble, whereas in Hydra A, the northern bubble still seems to be driving it. This is one of the main difficulties in making a realistic model for the shape of the shock front. In the simulation, the southern bubble is also somewhat further away from the shock than observations suggest. 

\item The simulation shows an offset temperature dip at approximately $(0,-100,0)$ kpc, which is not observed. The temperature contrast however is not very pronounced and may be washed out by projection. For a lower bulk flow velocity, this feature is absent (see Fig.~\ref{fig:slices_altern}). At that location, a surface brightness enhancement, also not present in the observation, can be seen in the projected X-ray map.
\end{itemize}

\subsection{Effects of AGN activity on the temperature structure in the cluster core}\label{sect:multit}

\begin{figure}
\includegraphics[width=\columnwidth]{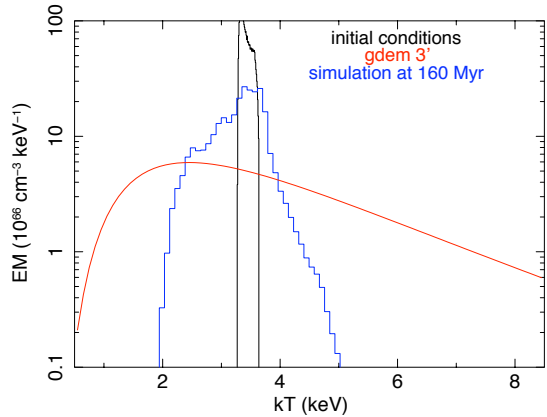}
\caption{Emission measure vs. temperature in a cylinder of 3\arcmin\ radius whose axis goes through the cluster center. In black, the initial conditions for the simulation are shown. In blue, the multi-temperature structure at the end-stage of the simulation, i.e. after 160 Myr, is plotted. The AGN activity significantly broadens the distribution compared to the initial conditions, but the best-fit Gaussian emission measure distribution model for the integrated spectrum in the central 3\arcmin\ region, shown in red, has an even greater full-width at half maximum.}
\label{fig:dydtsim}
\end{figure}

The plasma in many cool cores of galaxy clusters was proved to have an intrinsic multi-temperature structure \citep[e.g.][]{kaastra2004}. Among many examples, the most remarkable are M87 \citep{Belsole01,Molendi02,simionescu2007b}, Perseus \citep{sanders2007}, Centaurus \citep{matsushita2007b,sanders2008}, 2A0335+096 \citep{werner2006}. This multi-temperature structure can be due either to a small amount of gas which does cool radiatively down to low temperatures ($<1$keV) or to AGN-ICM interaction. The latter generates both hotter gas compared to the ambient (through shocks) and cooler gas which adiabatically expands as it is entrained by the rising radio lobes. Both of these processes are also observed in Hydra A \citep[][and this work]{Nulsen05,Wise07,Simionescu_HydraAI}. The temperature structure in Hydra A in fact is best described by a very broad Gaussian distribution with a full-width at half maximum of around 4.2~keV \citep{Simionescu_HydraAI}. We aim to determine from this simulation what is the effect of the AGN-ICM interaction on the multi-temperature structure and compare this to the multi-temperature structure observed in Hydra~A. To this end, we considered a cylinder with 3\arcmin\ radius going through the center of the simulation box along the z axis and constructed a histogram of the emission measure as a function of temperature at the initial and final stages of the simulation. Initially, only different temperatures as a function of radius (inside the central 3\arcmin\ or projected along the cylinder) contribute to the multi-temperature structure, which has a very narrow distribution. At the end of the simulation, the AGN activity has broadened the multi-temperature distribution significantly both towards higher and lower temperatures. However, we still do not achieve the best-fit width obtained for the temperature distribution in the central 3\arcmin\ spectrum extracted from the observation (Fig. \ref{fig:dydtsim}). This suggests either that the initial distribution of temperatures should be much broader than we assumed because of a more complex cluster formation history or that the history of the AGN outburst itself is more complex. A series of several outbursts following each other could for example contribute to a further broadening of the temperature distribution. Moreover, the cool core could have been more pronounced than we considered in the initial conditions of the simulation and could have been disrupted by the outburst, which could also contribute to a broadened temperature distribution towards the low temperature side. The shock could furthermore accelerate electrons and generate a non-Maxwellian tail in the electron distribution, which would mimic the presence of additional hot gas.  

\subsection{Alternative scenarios}
A physical set-up like the one described by the model, may prove very difficult to achieve in reality. The flow velocity required to reproduce the observed offset of the shock ellipse from the cluster center is large (about Mach 0.7) and must operate on a large spatial scale to effectively deflect the shock. It is not necessary in principle for the velocity field to be completely coherent - having a complex velocity field with an average velocity of that magnitude and orientation over this large spatial scale would suffice - but it is unclear whether accretion flows or cluster formation history can reproduce these conditions, even on average. 

Another option is to consider that the AGN and the central galaxy, rather than the rest of the ICM, are moving at $670\Kms$. However, peculiar velocities of BCGs are typically around only 160 $\Kms$ \citep{oegerle2001}. Although extreme examples have been observed, such as A 2255 showing two BCGs separated by as much as 2600 $\Kms$ \citep{Burns95} due to an ongoing merger, it is unlikely that the cD in Hydra A has such a high average velocity, especially given its current proximity to the cluster center.

An alternative scenario we simulated was that the large-scale velocity field is due to so-called ``sloshing'' of the ICM following, for example, a minor merger. We mimicked this in a simple manner by offsetting the whole ICM from the gravitational potential by 100 kpc. The ICM falls back into the potential well, leading to a large-scale bulk flow, which is, however non-homogeneous in space and time. Typical velocities are between $660 \Kms$ in the cluster center and $300 \Kms$ at 300 kpc. A temperature slice of this simulation taken at $160\Myr$ after the jet was launched is shown in Fig.~\ref{fig:slices_altern}. The ellipticity, energy and age of the shock remain similar, the offset is slightly larger than observed (93 kpc, which could be easily corrected by adjusting the initial offset or viewing direction) and, in this scenario, we begin to see an angle between the jets due to the fact that, after turning off the jet, its former channels are transported along with the flow. It is possible that the bend in the observed large-scale N radio lobe coincides with the bend in Fig. \ref{fig:slices_altern} and a more recent outburst in the cluster center along the NE-SW direction completed the radio morphology observed today. However, the sloshing seems to also produce an offset temperature dip which is not observed and the shock Mach number to the W is around 1.7, which is too high compared to the observed value (the Mach numbers in all other directions are consistent with the observed values).

Other possibilities to explain the shock shape in Hydra A include a combination of a smaller flow velocity in the cluster (at a more plausible value of $260\Kms$, the offset of the shock ellipse would be $28\Kpc$ and no additional substructure such as offset temperature dips would appear, Fig. \ref{fig:slices_altern}), asymmetrical density/temperature profiles such that shocks may run down different
profiles with different speeds, and a more pronounced influence of the bubbles in actively driving the shock front towards the north. The latter could be achieved, for example, if an existing older bubble recommences expansion due to a more recent outburst. 

The actual shape of the shock front, which is only roughly an ellipse (Sect. \ref{sect:geom}), and the radio morphology betray the likely necessity to include both a very complex flow pattern and a combination of additional physics and initial conditions before a truly realistic model is obtained. A detailed analysis of the cavity properties by \citet{Wise07}, as well as the multi-temperature structure in Hydra A (Sect. \ref{sect:multit}) both suggest a much more complex outburst history.
The simulation presented here is successful at reproducing several of the observed features and gives us insights about the next steps to be taken in order to further improve the modeling.

\section{Summary and Outlook}

We analyzed a deep XMM-Newton observation of the cluster of galaxies Hydra~A and focused on the large-scale shock discovered as a surface brightness discontinuity in Chandra images \citep{Nulsen05}. We find

\begin{itemize}

\item that the shock front can be seen in the pressure map as a ~20\% enhancement with respect to the radial average.

\item that the shape of the shock seen in the pressure map can be approximated with an ellipse with a semi-major axis of 360 kpc oriented 10 degrees clockwise from the N--S direction, a semi-minor axis of 275 kpc, and centered $\sim$70 kpc towards the NE with respect to the cluster center. This is a good simple approximation to the shock shape seen in the Chandra image, which shows however some more complex deviations from ellipticity.

\item for the first time, indications of temperature jumps corresponding to the shocked regions. We divided the data in four sectors towards the N, E, S and W and find temperature jumps with typical significances of 2$\sigma$. Combining the significances in the individual sectors we obtain a total significance of $4.3\sigma$.
\end{itemize}

We then used a spherically symmetric hydrodynamic model of a point explosion at the center of an initially isothermal, hydrostatic atmosphere \citep{Nulsen05} to simulate surface brightness profiles and temperature jumps across the shock. These were compared to the observational data to estimate the shock properties, such as Mach number, energy and age. We find

\begin{itemize}

\item that the Mach numbers determined from the temperature jumps in the shocked regions are in good agreement with the Mach numbers derived from EPIC/pn surface brightness profiles and previously from Chandra data \citep{Nulsen05}. This confirms that the large-scale surface brightness discontinuity in Hydra~A is due to a classical shock.

\item that the shock in all the four sectors has a Mach number consistent with $\sim$1.3, although the distance between the shock front and the cluster center differs. This is contrary to what we would expect from a point explosion.

\item estimated shock ages between 130 and 230 Myr. The larger shock age in the sectors where the shock is further from the cluster center suggests again that the shock generation mechanism is more complex. 

\item estimated shock energies between 1.5 and 3$\times 10^{61}$ ergs.

\end{itemize} 

To further improve the modeling of the shock in Hydra~A, we also employed 3D hydrodynamical simulations in which the shock is produced by a symmetrical pair of jets that originate from the cluster center, mimicking AGN activity. This creates an approximately elliptical shock front. To reproduce the observed 70 kpc offset between the cluster center and the center of the ellipse approximation to the shock front, we included large-scale coherent motions in the simulated ICM. We find

\begin{itemize}

\item that the simulation can successfully reproduce the size, ellipticity and average Mach number of the observed shock front.

\item that the variation of the Mach number along the simulated shock is small, although the shock is asymmetric. This is in good agreement with the observed properties and could not be explained with a simple point explosion model.

\item that the shock age and energy from the 3D simulation are 160 Myr and 3$\times 10^{61}$ ergs, respectively, within the range of the estimated values based on the 1D shock model.

\item that for the case of a potential flow around the central 100 kpc, the flow velocity needed to reproduce the observed offset of the center of the shock ellipse with respect to the cluster center is very high, 670 $\Kms$. 

\item that the AGN activity significantly broadens the temperature distribution in the cluster core.

\end{itemize}

However, such a high bulk flow velocity coherent over large regions in the ICM is unlikely, and the simulation does not reproduce the proximity of the observed northern radio lobe to the shock front, which is potentially an important additional factor contributing to the offset of the shock ellipse. The morphology of the radio lobes, especially the bending of the southern lobe, is also difficult to reproduce, suggesting the necessity for more detailed simulations. In an upcoming paper, we plan to further investigate the shock modeling using more complex initial conditions, such as an elliptical cluster potential and the presence of older bubbles through which the shock propagates. We will also vary the jet physics to include intermittent activity. Switching the jet on and off with different frequencies could put more momentum into bubbles, so that the shock becomes detached from them at a later stage. Moreover, the physics of the bubbles is important: we will check the effect of including sub-grid turbulence models which should prevent the shredding of the bubbles \citep[see][]{scannapieco08}, making them easier to bend and be affected by bulk flow motions. This should significantly alleviate the problems described above and provide a more realistic model.

\begin{acknowledgements}
We acknowledge the support by the DFG grant BR 2026/3 within the Priority Programme ``Witnesses of Cosmic History'', the supercomputing grants NIC
2195, 2256 and 2877 at the John-Neumann Institut at the Forschungszentrum J\"ulich, and NASA grant NNX07AQ18G 16610022. AS would like to thank the Harvard-Smithsonian CfA and Jacobs University Bremen for their hospitality. AF acknowledges support from BMBF/DLR under grant 50 OR 0207 and MPG.
This work is based on observations obtained with XMM-Newton, an ESA science mission with instruments and contributions directly funded by ESA member states and the USA (NASA). The Netherlands Institute for Space Research (SRON) is supported financially by NWO, the Netherlands Organization for Scientific Research. Part of the results presented were produced using the FLASH code, a product of the DOE
ASC/Alliances-funded Center for Astrophysical Thermonuclear Flashes at the University of Chicago.
\end{acknowledgements}

\bibliographystyle{aa}
\bibliography{clusters,bibliography,elke}

\end{document}